%% file: bare_conf_NDSS2021.tex
\documentclass[conference]{IEEEtran}
\ifCLASSINFOpdf
\else
\fi
\usepackage{bm}
\usepackage{multirow}
\usepackage{epsfig}
\usepackage{epstopdf}
\usepackage{url}
\usepackage{subfigure}
\usepackage{amsfonts}
\usepackage{xcolor}
\newcommand{\myfrac}
    [2]{\begin{array}{@{}c@{}}#1 \\[-0.75ex]#2\end{array}}

\begin{document}
%
\title{GALA: \underline{G}reedy Comput\underline{A}tion for \underline{L}inear \underline{A}lgebra in Privacy-Preserved Neural Networks}

\author{\IEEEauthorblockN{Qiao Zhang, Chunsheng Xin, and Hongyi Wu}
\IEEEauthorblockA{
Old Dominion University, Norfolk, VA 23529, USA\\
Email: qzhan002@odu.edu, cxin@odu.edu, h1wu@odu.edu}
}


%


\IEEEoverridecommandlockouts
\makeatletter\def\@IEEEpubidpullup{6.5\baselineskip}\makeatother
\IEEEpubid{\parbox{\columnwidth}{
    Network and Distributed Systems Security (NDSS) Symposium 2021\\
    21-24 February 2021\\
    ISBN 1-891562-66-5\\
    https://dx.doi.org/10.14722/ndss.2021.24351\\
    www.ndss-symposium.org
}
\hspace{\columnsep}\makebox[\columnwidth]{}}

\maketitle

\begin{abstract}
Machine Learning as a Service (MLaaS) is enabling a wide range of smart applications on end devices. However, privacy still remains a fundamental challenge. The schemes that exploit Homomorphic Encryption (HE)-based linear computations and Garbled Circuit (GC)-based nonlinear computations have demonstrated superior performance to enable privacy-preserved MLaaS. Nevertheless, there is still a significant gap in the computation speed. Our investigation has found that the HE-based linear computation  dominates the total computation time for  state-of-the-art deep neural networks. Furthermore, the most time-consuming component of the HE-based linear computation is a series of Permutation (Perm) operations that are imperative for dot product and convolution in privacy-preserved MLaaS. This work focuses on a deep optimization of the HE-based linear computations  to minimize the Perm operations, thus substantially reducing the overall computation time. To this end, we
propose {\em GALA: \underline{G}reedy comput\underline{A}tion for \underline{L}inear \underline{A}lgebra} in privacy-preserved neural networks, which views the HE-based linear computation as a series of Homomorphic Add, Mult and Perm operations and chooses the least expensive operation in each linear computation step to reduce the overall cost. GALA makes the following contributions: (1) It introduces a row-wise weight matrix encoding and combines the share generation
that is needed for the GC-based nonlinear computation, to reduce the Perm operations for the dot product; (2) It designs a first-Add-second-Perm approach (named {\em kernel grouping}) to reduce  Perm operations for convolution.
As such, GALA efficiently reduces the cost for the HE-based linear computation, which is a critical building block in almost all of the recent frameworks for privacy-preserved neural networks, including GAZELLE (Usenix Security'18), DELPHI (Usenix Security'20), and CrypTFlow2 (CCS'20). With its deep optimization of the HE-based linear computation, GALA can be a plug-and-play module integrated into these systems to further boost their efficiency. Our experiments show that it achieves a significant speedup up to 700$\times$ for the dot product and 14$\times$ for the convolution computation under different data dimensions.
Meanwhile, GALA demonstrates an encouraging runtime boost by 2.5$\times$, 2.7$\times$, 3.2$\times$, 8.3$\times$, 7.7$\times$, and 7.5$\times$ over GAZELLE
and 6.5$\times$, 6$\times$, 5.7$\times$, 4.5$\times$, 4.2$\times$, and 4.1$\times$ over CrypTFlow2, on AlexNet, VGG, ResNet-18, ResNet-50, ResNet-101, and ResNet-152, respectively.
\end{abstract}


%

\input{introduction}

\input{preliminaries}
\input{system_description_SecA}

\input{system_description_SecB}

\input{system_description_SecC}

\input{system_description_SecD}

\input{evaluation}

\input{conclusion}

\section*{Acknowledgment}
{\color{black}The authors would like to thank the shepherd and anonymous reviewers for the constructive and insightful guidance and comments. This work was supported in part by the National Science Foundation under Grant CNS-1828593, OAC-1829771, EEC-1840458, and CNS-1950704, Office of Naval Research under Grant N00014-20-1-2065, and the Commonwealth Cyber Initiative, an investment in the advancement of cyber R\&D, innovation and workforce development. For more information about CCI, visit cyberinitiative.org.}

\bibliographystyle{IEEEtranS}
\bibliography{ndss-sample}



%
%
%

\end{document}

%% file: introduction.tex
\section{Introduction}
Deep Learning (DL) is becoming prevalent and pervasive,
e.g., for pattern recognition~\cite{li2015convolutional}, medical diagnosis \cite{fakoor2013using}, speech recognition~\cite{dahl2012context} and credit-risk assessment~\cite{fan2018denoising}. In particular,  Convolutional Neural Network (CNN) has demonstrated superior performance in computer vision such as image classification~\cite{krizhevsky2012imagenet,simonyan2014very} and facial recognition~\cite{schroff2015facenet}. Since designing and training a deep neural network model requires intensive resource and DL talents, cloud providers begin to offer Machine Learning as a Service (MLaaS) \cite{wang2018rafiki}, where a proprietary DL model is trained and hosted on a cloud. Clients can utilize the service by simply sending queries (inference) to the cloud and receiving  results through a web portal. While this emerging cloud service is embraced as an important tool for efficiency and productivity, the interaction between clients and cloud servers leads to new vulnerabilities. This work focuses on the development of privacy-preserved and computationally efficient MLaaS.

Although communication can be readily secured from end to end, privacy still remains a fundamental challenge. On the one hand,  the clients must submit their data to the cloud for inference, but they want the data privacy well protected, preventing curious cloud provider or attacker with access to the cloud from mining valuable information. In many domains such as health care~\cite{mozaffari2015systematic} and finance~\cite{sohangir2018big}, data are extremely sensitive. For example, when patients transmit their physiological data to the server for medical diagnosis, they do not want anyone (including the cloud provider) to see it. Regulations such as Health Insurance Portability and Accountability Act (HIPAA) \cite{act1996health} and the recent General Data Protection Regulation (GDPR) in Europe~\cite{file2012proposal} have been in place to impose restrictions on sharing sensitive user information. On the other hand, cloud providers do not want users to be able to extract their proprietary model that has been trained with significant resource and efforts~\cite{tramer2016stealing}. Furthermore, the trained model contains private information about the training data set and can be exploited by malicious users~\cite{shokri2017membership,song2017machine,wang2018stealing}. To this end, there is an urgent need to develop effective and efficient schemes to ensure that, in MLaaS, a cloud server does not have access to users' data and a user cannot learn the server's model.

A series of efforts have been made to enable privacy-preserved MLaaS, by leveraging cryptographic techniques as summarized below. The first is the {\em Homomorphic Encryption (HE)-Based Approaches}. For example, in CryptoNets \cite{cryptonets}, Faster CryptoNets \cite{chou2018faster}
and CryptoDL \cite{hesamifard2018privacy}, the client encrypts data  using HE and sends the encrypted data to the server. The server performs polynomial computations (e.g., addition and multiplication) over encrypted data to calculate an encrypted inference result. The client finally obtains the inference outcome after decryption. E2DM \cite{jiang2018secure} adopts a more efficient HE (i.e., packed HE \cite{brakerski2013packed}) which packs multiple messages into one ciphertext and thus improves the computation efficiency. The second approaches is based on {\em Garbled Circuit (GC)}~\cite{yao1986generate}. DeepSecure \cite{rouhani2018deepsecure} and XONN \cite{riazi2019xonn} binarize the computations in neural networks and employ GC to obliviously obtain the prediction without leaking sensitive client data. The third approach exploits {\em Secret Sharing (SS)}. SS is used in \cite{wan2007privacy} and \cite{wagh2019securenn} to split the client data into shares. The server only owns one share of the data. The computations are completed by interactive share exchanges. In addition, Differential Privacy (DP) \cite{shokri2015privacy,abadi2016deep,phan2016differential} and Secure Enclave (SE)~\cite{mckeen2013sgx,ohrimenko2016oblivious,bayerl2020offline,zheng2017opaque} are also explored to protect data security and privacy in neural networks.
{\color{black}
In order to deal with  different properties of linearity (weighted sum and convolution functions) and nonlinearity (activation and pooling functions) in neural network computations, several efforts have been made to orchestrate multiple cryptographic techniques to achieve better performance \cite{zhang2018gelu, li2018falcon, juvekar2018gazelle, mohassel2017secureml, riazi2018chameleon, liu2017oblivious, zheng2019helen, chen2009privacy, yuan2013privacy, mohassel2018aby, xu2019cryptonn, chandran2017ezpc,boemer2020mp2ml, kumar2020cryptflow,rathee2020cryptflow2, mishra2020delphi}.}
{\color{black}
Among them, the schemes with HE-based linear computations and GC-based nonlinear computations (called the HE-GC neural network framework hereafter) demonstrate superior performance \cite{li2018falcon, juvekar2018gazelle, liu2017oblivious, mishra2020delphi}. Specifically, the GAZELLE framework \cite{juvekar2018gazelle} represents the state-of-the-art design for the HE-based linear computation and achieves a speedup of three orders of magnitude than the classic CryptoNets inference system~\cite{cryptonets}.
}

Despite the rapid improvement, there is still a significant gap in computation speed, rendering the existing schemes infeasible for practical applications.
{\color{black}
For example, the time constraints in many real-time applications (such as speech recognition) are within a few seconds~\cite{Alex2020,Google2020}.}
In contrast, our benchmark has shown that GAZELLE takes 43 seconds and 659 seconds to run the well-known deep neural networks ResNet-18 and ResNet-152~\cite{he2016deep} on an Intel i7-8700 3.2GHz CPU (see detailed experimental settings in Sec.~\ref{evaluation}), which renders it impractical in real-world applications.

This performance gap motivates us to further improve the efficiency of the HE-GC neural network frameworks. In  deep neural network, both the fully-connected and convolutional layers are based on the linear computation, while the activation functions perform nonlinear computation. The former dominates the total computation time in  state-of-the-art deep neural networks. For example, the runtime of the nonlinear computation in GAZELLE is merely 2.3\%, 1.8\%, 1.7\%, 1.5\%, 1.6\%, and 2\%, respectively, on AlexNet~\cite{krizhevsky2012imagenet}, VGG~\cite{simonyan2014very}, ResNet-18~\cite{he2016deep}, ResNet-50~\cite{he2016deep}, ResNet-101~\cite{he2016deep}, and  ResNet-152~\cite{he2016deep}. The nonlinear cost in the original plaintext models is even lower (averaged 1.7\%). This indicates a great potential to speed up the overall system through optimizing linear computations.
{\color{black}
Although a few recent approaches, e.g., DELPHI~\cite{mishra2020delphi} and CrypTFlow2~\cite{rathee2020cryptflow2}, perform better than GAZELLE in terms of the overall system runtime, they all inherit the HE-based linear computation in GAZELLE. This work contributes a solid optimization  on the HE-based linear computation  (i.e., dot product and convolution), which can be integrated into those systems (including  GAZELLE, DELPHI and CrypTFlow2) to further improve their overall system performance.}
The HE-based computation consists of three basic operations: Homomorphic Addition (Add), Multiplication (Mult), and Permutation (Perm). Our investigation has shown that the most time-consuming part of the HE-based computation is a series of Perm operations that are imperative to enable dot product and convolution. Our experiments show that Perm is 56 times slower than Add and 34 times slower than Mult. As shown in Table~\ref{cost_example}, in the dot product by multiplying a 2$\times$2048 matrix with a length-2048 vector, the cost in GAZELLE is dominated by Perm, which takes about \emph{98\% of the computation time}. This observation motivates the proposed linear optimization, which aims to minimize the Perm operations, thus substantially reducing the overall computation time. With less Perm operations, the proposed approach demonstrates 10$\times$ speedup in the above matrix-vector computation.

\begin{table}[!htbp]
\footnotesize
\centering
\caption{Cost of matrix-vector multiplication (time in millionsecond).}
\begin{tabular}{c|c|c|c|c|c|c|c}
\hline
\hline
\multirow{2}{*}{Method} & \multirow{2}{*}{Total (ms)} & \multicolumn{2}{c|}{Perm} & \multicolumn{2}{c|}{Mult} & \multicolumn{2}{c}{Add} \\
\cline{3-8}
 & & \# & time & \# & time & \# & time \\
\hline
GAZELLE & 2 & 11 & 1.96 & 2 & 0.01 & 11 & 0.037 \\
\hline
Proposed & 0.2 & 1 & 0.17 & 2 & 0.01 & 1 & 0.003 \\
\hline
\hline
\end{tabular}\label{cost_example}
\end{table}

This significant speedup lies in a simple and efficient idea to choose the least expensive operation in each linear computation step to reduce the overall cost. We name the proposed approach {\em GALA: \underline{G}reedy comput\underline{A}tion for \underline{L}inear \underline{A}lgebra} in privacy-preserved neural networks. We view the HE-based linear computation as a series of Homomorphic Add, Mult and Perm operations. The two inputs are the encrypted vector (or channels) from the client and the plaintext weight matrix (or kernel) from the server. The output is the encrypted dot product (or convolution). The objective in each step is to choose the most efficient operations in the descending priorities of Add, Mult and Perm. To this end, we (1) design a row-wise weight matrix encoding with combined share generation\footnote{The resultant linear output will be shared between server and client as the input of GC-based nonlinear computation.} (i.e., a row-encoding-share-RaS (Rotated and Sum) approach) to reduce the number of Perm operations in dot product by $\log_2{\frac{n}{n_o}}$ where $n$ is the number of slots in a ciphertext and $n_o$ is the output dimension of dot product, and (2) propose a first-Add-second-Perm approach (named {\em kernel grouping}) to reduce the number of Perm operations of convolution by a factor of $\frac{c_i}{c_n}$ where $c_i$ and $c_n$ are respectively the number of channels in input data and the number of channels that can be packed in a ciphertext. $n$ is always greater than and can be up to 8192 times of $n_o$ depending on the dimension of dataset~\cite{iris2020} and HE implementation~\cite{sealcrypto}.

{\color{black}
At the same time, $\frac{c_i}{c_n}$ is at least one and can be up to 256 for state-of-the-art neural network architectures such as ResNets~\cite{he2016deep} where the large channels, i.e., 1024 and 2048, and small kernels size, i.e., 1$\times$1 and 3$\times$3, are adopted. The larger input data from users will result in smaller $c_n$, which accordingly contributes to higher speedup especially in the state-of-the-art CNNs.
As such, GALA efficiently boosts the performance of HE-based linear computation, which is a critical building block in almost all of the recent frameworks for privacy-preserved neural networks, e.g., GAZELLE, DELPHI, and CrypTFlow2.
Furthermore, GALA's deep optimization of the HE-based linear computation can be integrated as a plug-and-play module into these systems to further improve their overall efficiency.
{\color{black}For example, GALA can serve as a computing module in the privacy-preserved DL platforms, MP2ML~\cite{boemer2020mp2ml} and CrypTFlow~\cite{kumar2020cryptflow}, which are compatible with the user-friendly TensorFlow~\cite{abadi2016tensorflow} DL framework.}
Our experiments show that GALA achieves a significant speedup up to 700$\times$ for the dot product and 14$\times$ for the convolution computation under various data dimensions.
Meanwhile, GALA demonstrates an encouraging runtime boost by 2.5$\times$, 2.7$\times$, 3.2$\times$, 8.3$\times$, 7.7$\times$, and 7.5$\times$ over GAZELLE
and 6.5$\times$, 6$\times$, 5.7$\times$, 4.5$\times$, 4.2$\times$, and 4.1$\times$ over CrypTFlow2, on AlexNet, VGG, ResNet-18, ResNet-50, ResNet-101, and ResNet-152, respectively. More details are given in Sec. \ref{evaluation}.
}

The rest of the paper is organized as follows. Sec. \ref{background} introduces the primitives that GALA relies on. Sec. \ref{sec:proposed} describes the design details of GALA. The experimental results are illustrated and discussed in Sec. \ref{evaluation}. Finally, Sec. \ref{conclusion} concludes the work.

%% file: preliminaries.tex
\section{Preliminaries}\label{background}
In this section, we introduce the overall system architecture and threat model, as well as cryptographic tools used in GALA.
\subsection{System Model}
\begin{figure}[!b]
\centering
\includegraphics[trim= {0cm 0cm 2cm 0cm}, clip, scale=0.5]{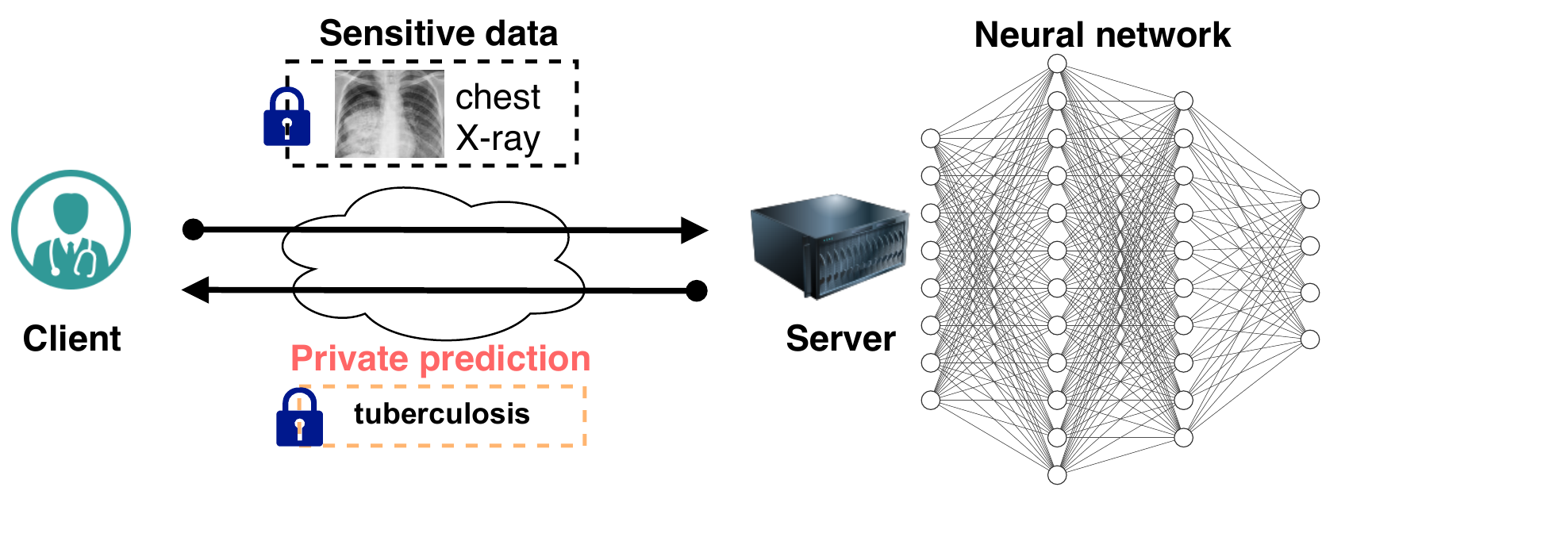}
\caption{An overview of the MLaaS system.}
\label{mlaas}
\end{figure}

We consider an MLaaS system shown in Fig.~\ref{mlaas}. The client  owns  private data. The server is in the cloud and has a well-trained deep learning model to provide the inference service based on the received  client's data. For example, a doctor sends an encrypted medical image (such as a chest X-ray) to the server, which runs the neural network model and returns the encrypted prediction to the doctor. The prediction is then decrypted into a plaintext result to assist diagnosis and health care planning.

\begin{figure}[!b]
\centering
\includegraphics[width=1\columnwidth]{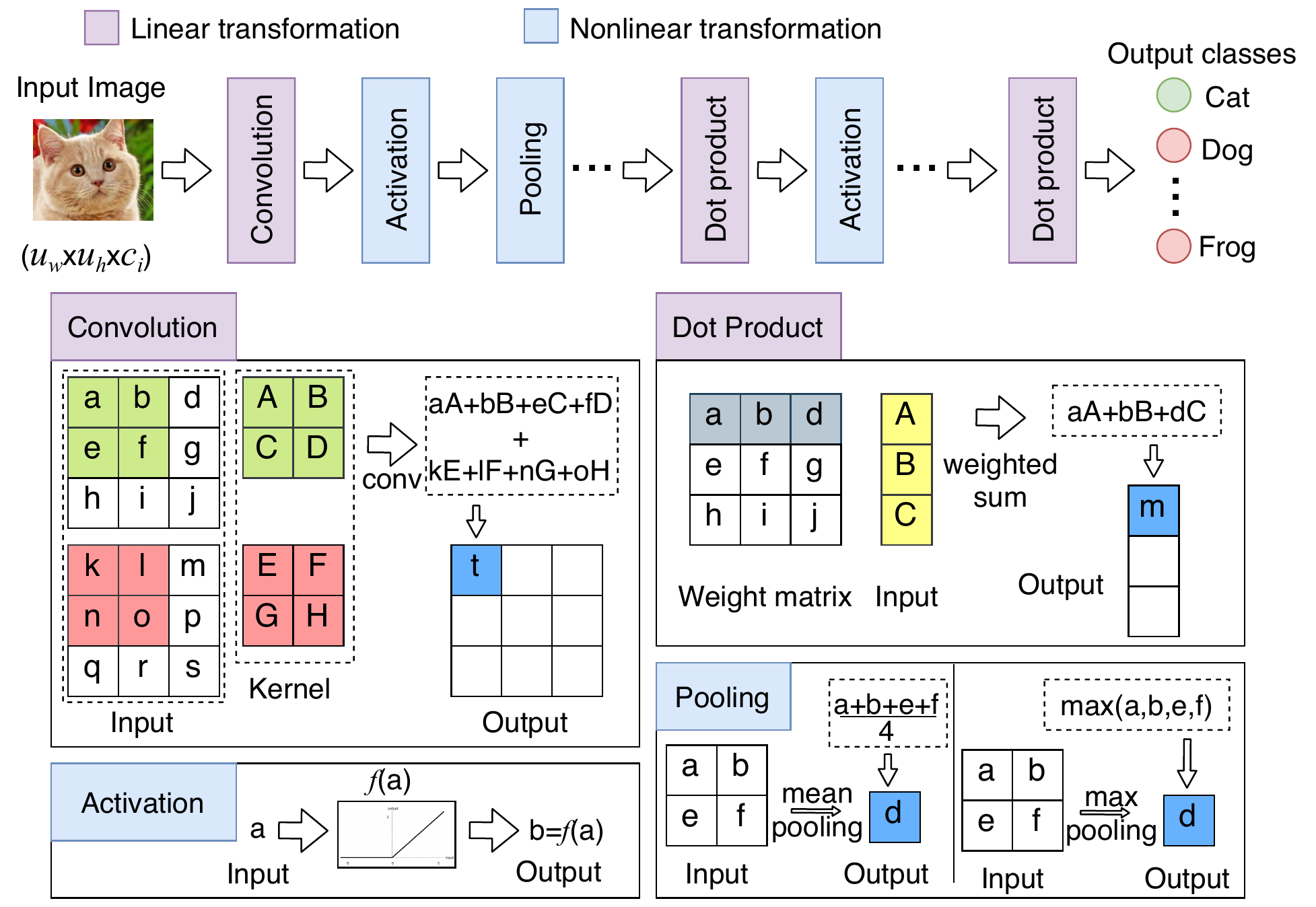}
\caption{An overview of CNN model.}
\label{overview}
\end{figure}

While various deep learning techniques can be employed to enable MLaaS, we focus on the Convolutional Neural Network (CNN), which has achieved wide success and  demonstrated superior performance in computer vision such as image classification~\cite{krizhevsky2012imagenet,simonyan2014very} and face recognition~\cite{schroff2015facenet}. A CNN consists of a stack of layers to learn a complex relation among the input data, e.g., the relations between pixels of an input image. It operates on a sequence of linear and nonlinear transformations to infer a result, e.g., whether an input medical image indicates that the patient has tuberculosis. The linear transformations are in two typical forms: \emph{dot product} (i.e., matrix-vector multiplication) and \emph{convolution}. The nonlinear transformations leverage \emph{activations} such as the Rectified Linear Unit (\emph{ReLu}) to approximate complex functions~\cite{hornik1991approximation} and \emph{pooling} (e.g., max pooling and mean pooling) for dimensionality reduction. CNN repeats the linear and nonlinear transformations recursively to reduce the high-dimensional input data to a low-dimensional feature vector for classification at the \emph{fully connected layer}. Without losing generality, we use image classification as an example in the following discussion, aiming to provide a lucid understanding of the CNN architecture as illustrated in Fig.~\ref{overview}.

\emph{Convolution.} The input to a convolutional layer has the dimension of $u_w\times u_h\times c_i$, where $u_w$ and $u_h$ are the width and height of the input feature map and $c_i$ is the number of the feature maps (or channels). For the first layer, the feature maps are simply the input image(s). Hereafter, we use the subscripts $i$ and $o$ to denote the input and  output, respectively. The input is convolved with $c_o$ groups of kernels. The size of each group of kernels is $k_w\times k_h\times c_i$, in which $k_w$ and $k_h$ are the width and height of the kernel. The number of channels of each kernel group must match with the input, i.e., $c_i$. The convolution will produce the feature output, with a size of $w_o\times h_o\times c_o$. Specifically, the process of convolution can be visualized as placing the kernel at different locations of the input data. At each location, a sum of element-wise product is computed between the kernel and corresponding data values within the kernel window, as shown in Fig.~\ref{overview}.

\emph{Dot Product.} The last convolutional layer is typically connected with the fully-connected layer, which computes the weighted sum, i.e., a dot product between the weight matrix $w$ of size $n_o\times n_i$ and a flatten feature vector of size $n_i\times 1$. The output is a vector with the size of $n_o\times 1$. Each element of the output vector is calculated as a sum of element-wise product between one row of weight matrix and the flatten feature vector, as shown in Fig.~\ref{overview}.

\emph{Activation.} Nonlinear activation is applied to convolutional and weighted-sum outputs in an elementwise manner, as shown in Fig.~\ref{overview}. The commonly used activation functions include \textit{ReLu}, $f(x)=\max\{0,x\}$; \emph{sigmoid}, $f(x)=\frac{1}{1+e^{-x}}$; and \emph{tanh}, $f(x)=\frac{e^{2x}-1}{e^{2x}+1}$. The last layer uses the \emph{softmax} function $f(x)=\frac{e^{x}}{\sum_ie^{x(i)}}$ to normalize the output into a probability vector.

\emph{Pooling.} Pooling conducts downsampling to reduce dimensionality. In this work, we consider \emph{Mean pooling}, which is implemented in CryptoNets and also commonly adopted in state-of-the-art CNNs. It splits a feature map into regions and averages the regional elements. Compared to max pooling (another pooling function which selects the maximum value in each region), authors in~\cite{zhou2016learning} have claimed that while the max and mean pooling functions are rather similar, the use of mean pooling encourages the network to identify the complete extent of the object,
which builds a generic localizable deep representation that exposes the implicit attention of CNNs on an image. In HE-GC neural network frameworks, the mean pooling is easily conducted on the shares of both client and server, without extra cost~\cite{liu2017oblivious,juvekar2018gazelle}.

In this work, we mainly focus on privacy-preserved linear optimization (i.e., convolution and dot product). The privacy-preserved nonlinear optimizations (especially activations) are based on GC as introduced in other HE-GC approaches such as GAZELLE \cite{juvekar2018gazelle}.

\subsection{Threat Model}
Similar to GAZELLE~\cite{juvekar2018gazelle} and other previous works, namely the SecureML~\cite{mohassel2017secureml}, MiniONN~\cite{liu2017oblivious},
DeepSecure~\cite{rouhani2018deepsecure} and XONN~\cite{riazi2019xonn},
we adopt the semi-honest model, in which both parties try to learn additional information from the message received (assuming they have a bounded computational capability). That is, the client $\mathcal{C}$ and server $\mathcal{S}$ will follow the protocol, but $\mathcal{C}$ wants to learn  model parameters and $\mathcal{S}$ attempts to learn the client's data. Note that, many applications are built on well-known deep network structures such as AlexNet~\cite{krizhevsky2012imagenet}, VGG-16/19~\cite{simonyan2014very} and ResNet-50~\cite{he2016deep}. Hence we do not intend to protect the structure (number of layers, kernel size, etc), but focus on the protection of model parameters. In the case that the implemented structure is proprietary and has to be protected, the server can introduce redundant layers and kernels to hide the real structure at a computational expense~\cite{liu2017oblivious, juvekar2018gazelle}. Hence, the overarching goal is to make the server oblivious of the private data from the client, and prevent the client from learning  model parameters of the server. GAZELLE has demonstrated the security of HE-GC neural network framework according to the cryptographic standard of ideal/real security~\cite{goldreich2019play, goldreich2009foundations, goldwasser1989knowledge}. The same security framework is adopted in this work.

{\color{black}
Note that, while the client can use the server's prediction service as a blackbox oracle to extract the model~\cite{tramer2016stealing,wang2018stealing}, or even infer the training set~\cite{fredrikson2015model, nasr2018comprehensive, shokri2017membership},
GALA does not aim to protect against the black-box attack. Instead, it focuses on protecting the input data and the model parameters during the inference process, which stays in line with the threat model of GAZELLE~\cite{juvekar2018gazelle}, SecureML~\cite{mohassel2017secureml}, DELPHI~\cite{mishra2020delphi}, CrytoFlow2~\cite{rathee2020cryptflow2}, etc., the output of neural network model is returned to the client which decrypts the result and gets the plaintext prediction.
}

\subsection{Cryptographic Tools}\label{priliminary:crypto}
The proposed privacy-preserved deep neural network framework, i.e., GALA, employs three fundamental cryptographic tools as outlined below.

\emph{(1) Packed Homomorphic Encryption}.
Homomorphic Encryption (HE) is a cryptographic primitive that supports meaningful computations on encrypted data
without the decryption key, which has found increasing applications in data communications, storage and computations~\cite{takabi2010security}. Traditional HE operates on individual ciphertext~\cite{paillier1999public}, while the \emph{packed homomorphic encryption} (PHE) enables packing of multiple values into a single ciphertext and performs component-wise homomorphic computation in a Single Instruction Multiple Data (SIMD) manner~\cite{brakerski2013packed} to take  advantage of parallelism. Among various PHE techniques, our work builds on the Brakerski-Fan-Vercauteren (BFV) scheme \cite{fan2012somewhat}, which involves four parameters\footnote{The readers are referred to \cite{juvekar2018gazelle} for more detail.}: 1) ciphertext modulus $q$, 2) plaintext modulus $p$, 3) the number of ciphertext slots $n$, and 4) a Gaussian noise with a standard deviation $\sigma$. The secure computation involves two parties, i.e., the client $\mathcal{C}$ and server $\mathcal{S}$.

In PHE, the encryption algorithm encrypts a plaintext message vector $x$ from $\mathbb{Z}^n$ into a ciphertext $[x]$ with $n$ slots. We denote $[x]_{\mathcal{C}}$ and $[x]_{\mathcal{S}}$ as the ciphertext encrypted by client $\mathcal{C}$ and server $\mathcal{S}$, respectively. The decryption algorithm returns the plaintext vector $x$ from the ciphertext $[x]$. Computation can be performed on the ciphertext. In a general sense, an evaluation algorithm inputs several ciphertext $[x_1],[x_2],\cdots,$ and outputs a ciphertext $[x']=f([x_1],[x_2],\cdots)$. The function $f$ is constructed by homomorphic addition (Add), Multiplication (Mult) and permutation (Perm). Specifically, Add($[x]$,$[y]$) outputs a ciphertext $[x+y]$ which encrypts the elementwise sum of $x$ and $y$.
Mult($[x]$,$s$) outputs a ciphertext $[x\odot{s}]$ which encrypts the elementwise multiplication of $x$ and plaintext $s$.
It is worth pointing out that GALA is designed to require scalar multiplication between a ciphertext and a plaintext only, but not the much more expensive multiplication between two ciphertext. Hereafter, we use ScMult to denote the scalar multiplication involved in GALA.
Perm($[x]$) permutes the $n$ elements in $[x]$ into another ciphertext $[x_\pi]$,  where $x_\pi=(x({\pi{_0}}),x({\pi{_1}}),\cdots)$ and $\pi_i$ is a permutation of  $\{0,1,\cdots,n-1\}$. Additionally, the computation cost for a series of Perm operations on the same ciphertext can be optimized by first conducting one Perm Decomposition (DecPerm) on the ciphertext and then doing the corresponding series of Hoisted Perm (HstPerm) operations~\cite{juvekar2018gazelle}. Since only one DecPerm is involved, it can amortize the total permutation time.

The run-time  of Add and ScMult is significantly lower than that of Perm. From our experiments, a Perm operation is 56 times slower than an Add operation and 34 times slower than a ScMult operation. This observation motivates the proposed linear optimization, which aims to minimize the number of  Perm operations, thus substantially reducing the overall computation time.

Meanwhile, PHE introduces noise in the ciphertext which theoretically hides the original message \cite{juvekar2018gazelle, brakerski2012fully}.
Assume the noise of $[x]$ and $[y]$ are $\eta_0$ and $\eta_1$, then the noise after the Add operation is approximately $\eta_0+\eta_1$. The noise after a ScMult operation is $\eta_{mult}\eta_0$ where $\eta_{mult}$ is the \textit{multiplicative noise growth of the SIMD scalar multiplication operation}~\cite{juvekar2018gazelle}. The noise after a Perm operation is $\eta_0+\eta_{rot}$ where $\eta_{rot}$ is the \textit{additive noise growth
of a permutation operation}~\cite{juvekar2018gazelle}.
Roughly, we have $\eta_{rot}>\eta_{mult}\gg\eta_0\gg1$.
If the noise goes beyond a certain level, the decryption would fail. Thus it is also important to have a good noise management over the ciphertext. We will show in Sec.~\ref{noise_manage} that GALA has a better noise control than GAZELLE, which further guarantees the overall success for the linear computations.

\emph{(2) Secret Sharing}. In the secret sharing protocol, a value is shared between two parties, such that combining the two secrets yields the true value \cite{riazi2018chameleon}. In order to additively share a secret $m$, a random number, $r$, is selected and two shares are created as $\langle{m}\rangle_0=r$ and $\langle{m}\rangle_1=m-r$. Here, $m$ can be either plaintext or ciphertext. A party that wants to share a secret sends one of the shares to the other party. To reconstruct the secret, one needs to only add two shares $m=\langle{m}\rangle_0+\langle{m}\rangle_1$.

While the overall idea of secret sharing (SS) is straightforward, creative designs are often required to enable its effective application in practice. Specifically, in the HE-GC neural network framework, the linear result from the dot product or convolution is encrypted at the server side and needs to be shared with the client to enable the following GC-based nonlinear computation.
Assume $m$ is the resulted ciphertext of a linear computation at the server, GAZELLE then generates the share $\langle{m}\rangle_0=r$ and sends $\langle{m}\rangle_1=m-r$ to the client. The two shares act as the input of the GC-based nonlinear computation. Here the computation of $m$ involves a series of Perm operations, which is time-consuming. Instead of directly generating the share $\langle{m}\rangle_0=r$ for $m$, we develop a \textit{share-RaS (Rotate and Sum) computing} for dot product which lets the server generate an indirect share $r'$ for the incomplete $m$, $m'$, while the true $r$ is easy to be derived from $r'$ and the true $\langle{m}\rangle_1=m-r$ is easy to be derived from $m'-r'$. The  computation of $m'$ eliminates a large number of Perm operations thus reducing the computation complexity. Specifically, Our result shows that the proposed \textit{share-RaS computing} demonstrates
19$\times$ speedup for the dot product by multiplying a 16$\times$128 matrix with a length-128 vector (the detailed benchmarks are shown in Sec. \ref{evaluation}).

{\color{black}
\emph{(3) Oblivious Transfer}. In the 1-out-of-$k$ Oblivious Transfer (OT)~\cite{brassard1986all}, denoted as $(\myfrac{k}{1})$-OT$_\ell$,
the sender's inputs are the $k$ strings, $m_0, m_1, \cdots, m_{k-1}\in{\{0, 1\}^{\ell}}$, and the receiver's input is a value $i\in{\{0, 1, \cdots, k-1\}}$.
At the end of the OT execution, the receiver obtains $m_i$ from the functionality
and the sender receives no output. Here, the OT
protocol guarantees that 1) the receiver learns nothing about
$m_{j,j\neq{i}}$, and 2) the sender learns nothing about $i$.
An advancement in the practicality of
OT protocols is the  OT extension~\cite{ishai2003extending}, which is further optimized such as in~\cite{kolesnikov2013improved}.
A special type of OT extension is the correlated OT extension (COT)~\cite{asharov2013more}.
Particularly, the 1-out-of-2 COT, denoted as $(\myfrac{2}{1})$-COT$_\ell$, can be used for linear computation\footnote{We refer readers to~\cite{beaver1991efficient, demmler2015aby, mohassel2017secureml, rathee2020cryptflow2} for more details.}.
In $(\myfrac{2}{1})$-COT$_\ell$, the sender's two inputs to each OT
are not independent. Instead, the two inputs to each OT instance
are a random value $s_0$ and a value $s_1 = f(s_0)$ for a correlation
function $f$ of the sender's choice. The receiver obtains either $s_0$ or $s_1$ as output
depending on $b$.

}

%% file: system_description_SecA.tex
\section{System Description}\label{sec:proposed}

In this section, we introduce the proposed system, GALA, for streamlining the linear computations (i.e., matrix-vector multiplication and convolution) in privacy-preserved neural network models. The HE-based linear computation consists of three basic operations: Homomorphic Addition (Add), Multiplication (Mult), and Permutation (Perm). Our investigation has shown that the linear computation dominates the total computation cost and the most time-consuming part of HE-based linear computation is a series of Perm operations that are imperative to enable dot product and convolution. GALA aims to minimize the Perm operations, thus substantially reducing the overall computation time.
We view the HE-based linear computation as a series of Add, Mult and Perm. The two inputs to linear computation are the encrypted vector (or channels) from the client and the plaintext weight matrix (or kernel) from the server. The output is the encrypted dot product (or convolution). The objective in each step is to choose the most efficient operations in the descending priorities of Add, Mult and Perm.
{\color{black}
Therefore, the overhead for the HE-based linear computation can be efficiently reduced by GALA. The recent privacy-preserved neural network frameworks can integrate GALA as a plug-and-play module to further boost their efficiency.}
We also analyze the (better) noise management and (guaranteed) system security of GALA.

\subsection{Row-encoding-share-RaS Matrix-Vector Multiplication}\label{sys:matrix_vector}
We first focus on matrix-vector multiplication (dot product) which multiplies a plaintext matrix at the server with an encrypted vector from the client. We first discuss a naive method followed by the mechanism employed in the state-of-the-art framework (i.e., GAZELLE \cite{juvekar2018gazelle}), and then introduce the proposed optimization of GALA
that significantly improves the efficiency in matrix-vector multiplication.

For a lucid presentation of the proposed GALA and comparison with the state-of-the-art framework, we adopt the same system model used in \cite{juvekar2018gazelle}. More specifically,  we consider a Fully Connected (FC) layer with $n_i$ inputs and $n_o$ outputs. The number of slots in one ciphertext is $n$. We also adopt the assumptions used in \cite{juvekar2018gazelle}: $n$, $n_i$ and $n_o$ are powers of two, and $n_o$ and $n_i$ are smaller than $n$. If they are larger than $n$, the original $n_o\times{n_i}$ matrix can be split into $n{\times}n$ sized blocks that are processed independently.

\begin{figure}[!bp]
\centering
\includegraphics[scale=0.31]{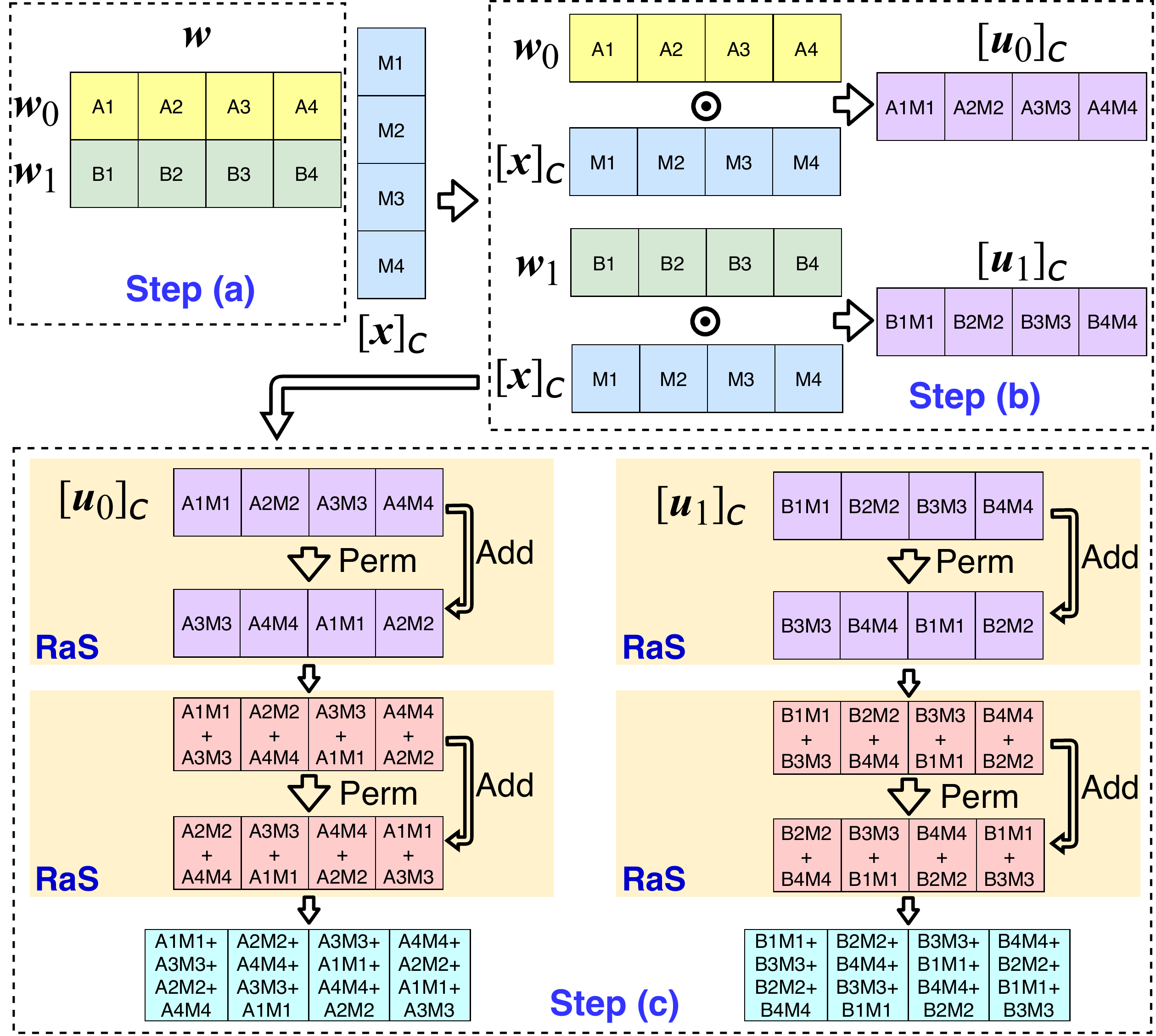}
\caption{Naive matrix-vector multiplication.}
\label{weight_sum_diagram1}
\end{figure}

\vspace*{0.05in}\noindent\textbf{1) Naive Method:}
The naive calculation for matrix-vector multiplication is shown in Figure \ref{weight_sum_diagram1}, where $\bm{w}$ is the $n_o\times{n_i}$ plaintext matrix on the server and $[\bm{x}]_{\mathcal{C}}$ is the HE-encrypted vector provided by the client. The server encodes each row of $\bm{w}$ into a separate plaintext vector (see step (a) in Figure \ref{weight_sum_diagram1}). The length of each encoded vector is $n$ (including padded 0's if necessary).
We denote these encoded plaintext vectors as $\bm{w}_0, \bm{w}_1, \cdots, \bm{w}_{(n_o-1)}$. For example, the yellow and green rows in step (a) of Figure \ref{weight_sum_diagram1} are $\bm{w}_0$ and $\bm{w}_1$, respectively.

The server intends to compute the dot product between $\bm{w}$ and $[\bm{x}]_{\mathcal{C}}$. To this end, it first uses ScMult to compute the elementwise multiplication between $\bm{w}_i$ and the encrypted input vector $[\bm{x}]_{\mathcal{C}}$ to get $[\bm{u}_i]_{\mathcal{C}} = [\bm{w}_i\odot\bm{x}]_{\mathcal{C}}$ (see step (b) in Figure \ref{weight_sum_diagram1}). The sum of all elements in $\bm{u}_i$ will be the $i$-th element of the desired dot product between $\bm{w}$ and $[\bm{x}]_{\mathcal{C}}$. However, as discussed in Sec. \ref{priliminary:crypto}, it is not straightforward to obtain the sum under the packed HE. A \emph{rotate-and-sum}  (RaS) calculation must be used here, as illustrated in step (c) of Figure \ref{weight_sum_diagram1}. Specifically, the entries in $[\bm{u}_i]_{\mathcal{C}}$ are first rotated through  Perm  by $\frac{n_i}{2}$ positions such that the first $\frac{n_i}{2}$ entries of the rotated $[\bm{u}_i]_{\mathcal{C}}$ are actually the second $\frac{n_i}{2}$ entries of the original $[\bm{u}_i]_{\mathcal{C}}$. 
Then the server uses Add to conduct elementwise addition between the rotated $[\bm{u}_i]_{\mathcal{C}}$ and the original $[\bm{u}_i]_{\mathcal{C}}$, which results in a ciphertext whose first $\frac{n_i}{2}$ entries contain the elementwise sum of the first and second $\frac{n_i}{2}$ entries of $\bm{u}_i$. The server conducts this RaS process for $\log_2n_i$ iterations. Each iteration acts on the resulted ciphertext from the previous iteration, and rotates by half of the previous positions, as shown in Step (c) of Figure \ref{weight_sum_diagram1}. Finally, the server gets a ciphertext where the first entry is the $i$-th element in $\bm{wx}$. By applying this procedure on each of the $n_o$ rows (i.e., $\bm{w}_0, \bm{w}_1, \cdots, \bm{w}_{(n_o-1)}$), the server obtains $n_o$ ciphertext. Altogether, the first entries of those ciphertext  correspond to $\bm{wx}$.

We now analyze the complexity of the above linear computation process, in terms of the number of operations and output ciphertext. We consider the process starting from the server's reception of $[\bm{x}]_{\mathcal{C}}$ (i.e., the encrypted input data from the client) until it obtains the to-be-shared ciphertext\footnote{In HE-GC neural network computing, the resultant ciphertext from linear calculation are shared between client and server as the input of GC-based nonlinear function.} (i.e., the $n_o$ ciphertext after RaS).
There are a total of $n_o$ scalar multiplications (ScMult) operations, $n_o\log_2n_i$  Perm operations and $n_o\log_2 n_i$ Add operations. It yields  $n_o$ output ciphertext, each of which contains one element of the linear result $\bm{wx}$. This inefficient use of the ciphertext space results in a low efficiency for linear computations.

\begin{figure}[!tbp]
\centering
\includegraphics[scale=0.3]{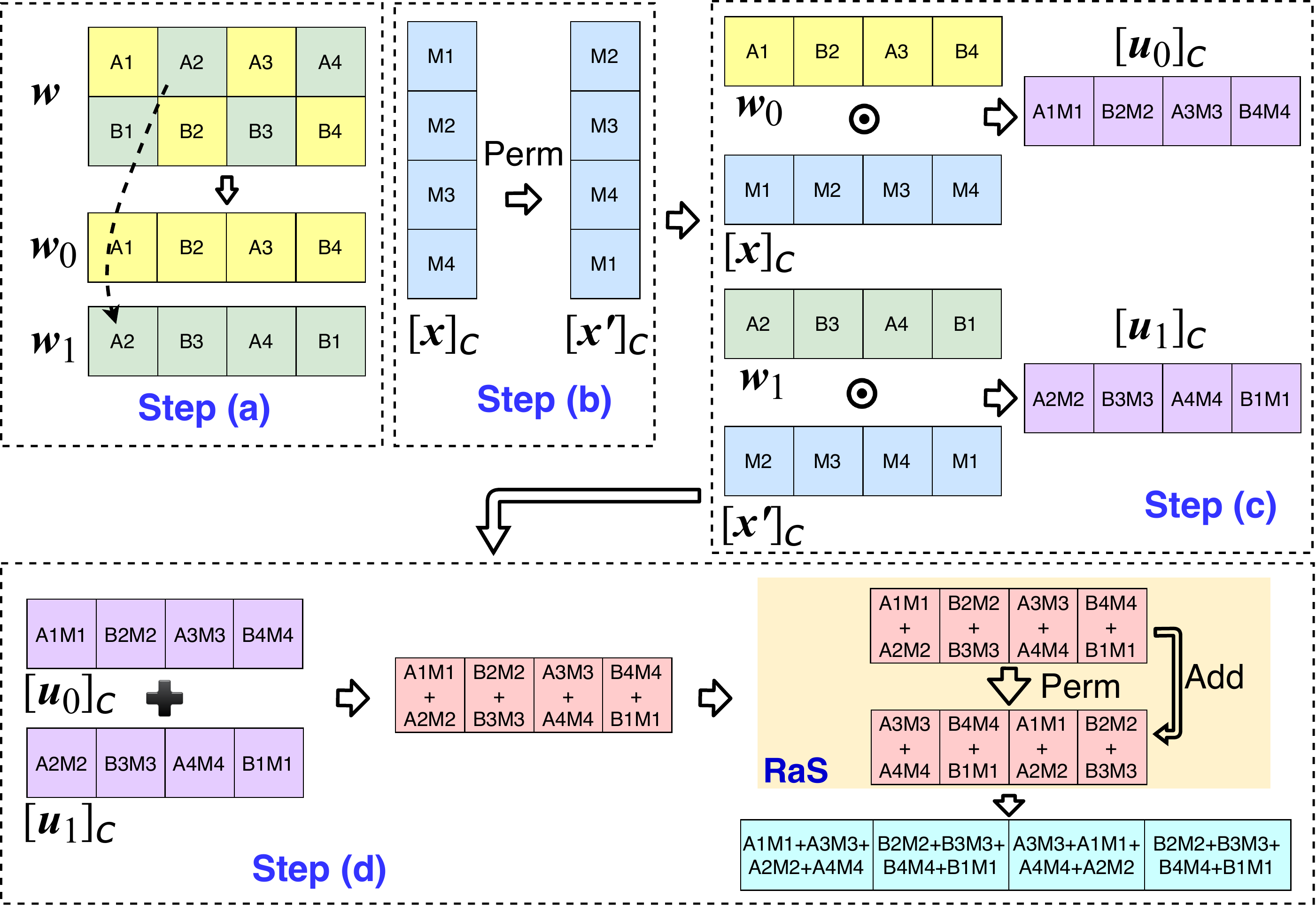}
\caption{Hybrid matrix-vector multiplication.}
\label{weight_sum_diagram2}
\end{figure}

\vspace*{0.05in}\noindent\textbf{2) Hybrid Calculation (GAZELLE):}
In order to fully utilize the $n$ slots in a ciphertext and further reduce the complexity, the state-of-the-art scheme is to combine the diagonal encoding \cite{halevi2014algorithms} and RaS, by leveraging the fact that $n_o$ is usually much smaller than $n_i$ in FC layers. This hybrid method shows that the number of  expensive  Perm operations is a function of $n_o$ rather than $n_i$, thus accelerating the computation of FC layers~\cite{juvekar2018gazelle}. The basic idea of the hybrid method is shown in Figure~\ref{weight_sum_diagram2}.

Specifically, the server encodes $\bm{w}$ into $n_o$ plaintext vectors through a diagonal manner.
For example, in step (a) of Figure~\ref{weight_sum_diagram2}, the first plaintext vector $\bm{w}_0$ consists of the yellow elements of matrix $\bm{w}$, (A1, B2, A3, B4), and the second plaintext vector $\bm{w}_1$ consists of the green elements (A2, B3, A4, B1). Note that the $\bm{w}_0$ in this method is different from the $\bm{w}_0$ in the naive method of Figure~\ref{weight_sum_diagram1}. So is $\bm{w}_1$.

The server then rotates $[\bm{x}]_{\mathcal{C}}$ by $i$ positions, shown in step (b), and uses ScMult to perform elementwise multiplication with $\bm{w}_i$. For example, in step (c) of Figure~\ref{weight_sum_diagram2}, $\bm{w}_0$  is multiplied with the encrypted data $[\bm{x}]_{\mathcal{C}}$ and $\bm{w}_1$  is multiplied with the input that is rotated by one position (i.e., $[\bm{x}']_{\mathcal{C}}$).
As a result, the server gets $n_o$ multiplied ciphertext, $\{[\bm{u}_i]_{\mathcal{C}}\}$. The entries in each of $\{[\bm{u}_i]_{\mathcal{C}}\}$  are  partial sums of the elements in the matrix-vector multiplication  $\bm{wx}$. For example, as shown in step (c) of Figure \ref{weight_sum_diagram2}, the server obtains two multiplied ciphertext (i.e., $[\bm{u}_0]_{\mathcal{C}}$ and $[\bm{u}_1]_{\mathcal{C}}$) whose elements are  partial sums of the first and second elements of $\bm{wx}$ (i.e., (A1M1 + A2M2 + A3M3 + A4M4) and (B1M1 + B2M2 + B3M3 + B4M4)). Then the server sums them up elementwise, to form another ciphertext, which is the vector in the middle of step (d) in Figure~\ref{weight_sum_diagram2}.
At this point, similar to the naive method, the server proceeds with $\log_2{\frac{n_i}{n_o}}$ RaS iterations and finally obtains a single ciphertext whose first $n_o$ entries are the corresponding $n_o$ elements of $\bm{wx}$ (see the first two elements of the vector after RaS in step (d)).

Furthermore, as the number of slots $n$ in a ciphertext is always larger than the dimension of the input vector, $n_i$, the computation cost is further reduced by packing copies of input $\bm{x}$ as much as possible to form $[\bm{x}_\textmd{pack}]_{\mathcal{C}}$. Thus $[\bm{x}_\textmd{pack}]_{\mathcal{C}}$ has $\frac{n}{n_i}$ copies of $\bm{x}$ and the server is able to multiply $\frac{n}{n_i}$ encoded vectors with $[\bm{x}_{\textmd{pack}}]_{\mathcal{C}}$ by one ScMult operation. Therefore the server gets $\frac{n_in_o}{n}$ rather than $n_o$ multiplied ciphertext. The resulted single ciphertext now has $\frac{n}{n_o}$ rather than $\frac{n_i}{n_o}$ blocks. The server then applies $\log_2{\frac{n}{n_o}}$ RaS iterations to get the final ciphertext, whose first $n_o$ entries are the $n_o$ elements of $\bm{wx}$.

The hybrid method requires $\frac{n_in_o}{n}$ scalar multiplications (ScMult), $\frac{n_in_o}{n}-1$ HstPerm rotations for $[\bm{x}_{\textmd{pack}}]_{\mathcal{C}}$, $\log_2\frac{n}{n_o}$ Perm rotations, and $\frac{n_in_o}{n}+\log_2\frac{n}{n_o}-1$ additions (Add). There is only one output ciphertext, which efficiently improves the slot utilization compared to the naive method.

\vspace*{0.05in}\noindent\textbf{3) Row-encoding-share-RaS Multiplication (GALA):}
The proposed GALA framework is motivated by two observations on the hybrid method. First, the hybrid method essentially strikes a tradeoff between Perm and HstPerm operations, where the number of Perms (which is the most expensive HE operation) is proportional to the number of slots in a ciphertext. This is not desired as we prefer a large $n$ to pack more data for efficient SIMD HE. GALA aims to make the number of Perm operations disproportional to the number of slots and eliminate all HstPerm operations on the input ciphertext.

The second observation is  the $\log_2\frac{n}{n_o}$ RaS operations. We discover that this is actually unnecessary. Specifically, the unique feature in the HE-GC neural network framework is that the resultant single ciphertext from  linear computing is shared between the client and server, to be the input for the nonlinear computing in the next phase. As the shares are in plaintext, we propose to transfer the final $\log_2\frac{n}{n_o}$ RaS operations in the HE domain to $\log_2\frac{n}{n_o}$ RaS operations in plaintext. This significantly  reduces expensive Perm operations.
For example, multiplying a 16$\times$128 matrix with a length-128 vector by our proposed scheme shows about 19$\times$ speedup compared with the hybrid method~\cite{juvekar2018gazelle} on a commodity machine (see detailed benchmarks in Sec. \ref{evaluation}).

\begin{figure}[!tbp]
\centering
\includegraphics[trim= {0cm 0cm 0cm 0cm}, clip, scale=0.308]{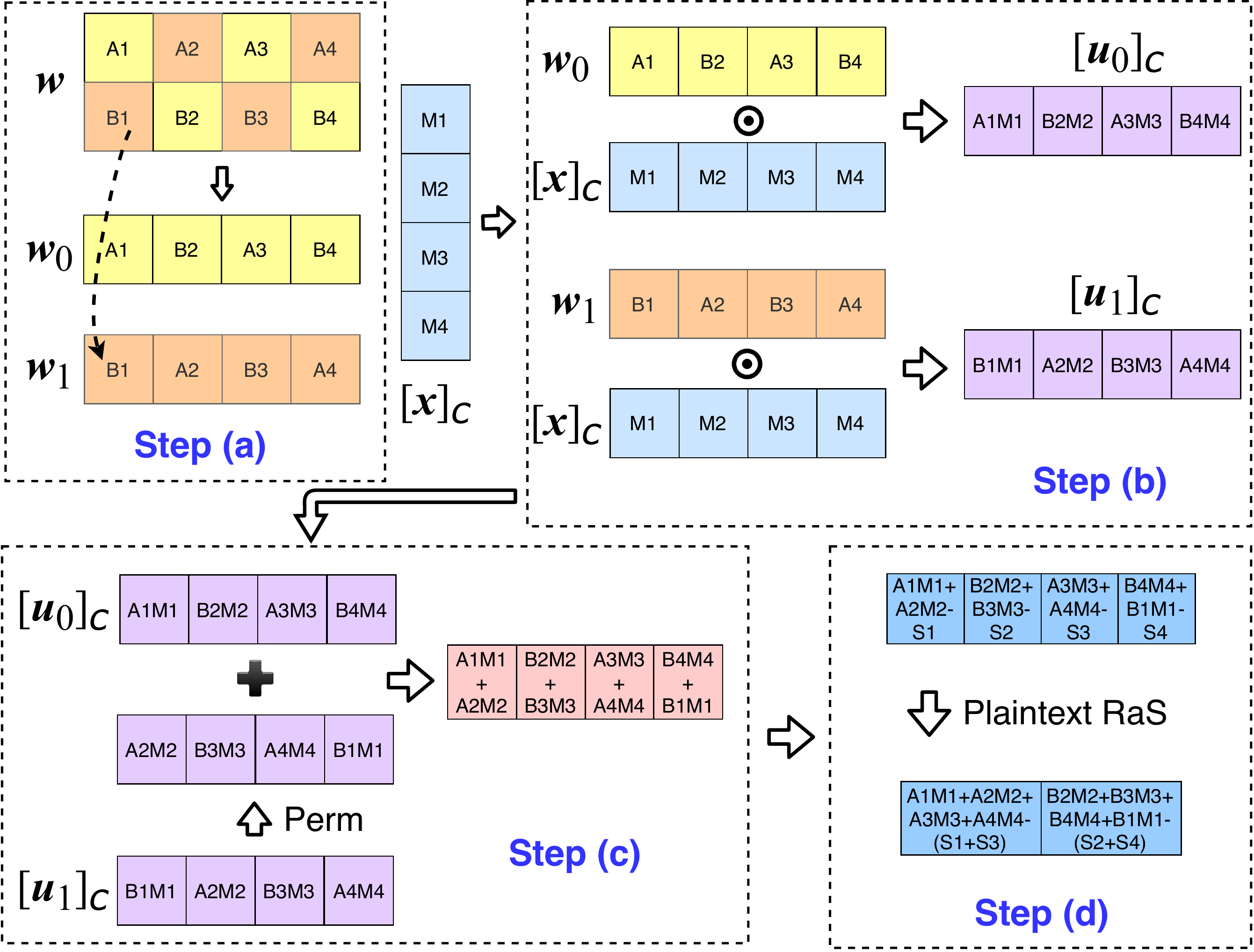}
\caption{Row-encoding-share-RaS multiplication.}
\label{weight_sum_diagram3}
\end{figure}

Figure \ref{weight_sum_diagram3} illustrates GALA's matrix-vector calculation. The server first conducts the row-wise weight matrix encoding which  encodes $\bm{w}$ into $n_o$ plaintext vectors in a diagonal manner,  as shown in step (a) in Figure~\ref{weight_sum_diagram3}. Compared with the hybrid method, the row-wise weight matrix encoding of GALA enables the server to directly multiply $\bm{w}_i$ and $[\bm{x}]_{\mathcal{C}}$, eliminating the Perm operations on $[\bm{x}]_{\mathcal{C}}$ in  step (b). Furthermore, the encoding also benefits the noise management in the resultant to-be-shared ciphertext as to be analyzed in Sec. \ref{noise_manage}.

As a result, the server gets $n_o$ multiplied ciphertext, $\{[\bm{u}_i]_{\mathcal{C}}\}$, such that the first entry of $[\bm{u}_i]_{\mathcal{C}}$ is a partial sum of the $i$-th element of the matrix-vector multiplication $\bm{wx}$. For example, in step (b) of Figure~\ref{weight_sum_diagram3}, the first element A1M1  in $[\bm{u}_0]_{\mathcal{C}}$ is a partial sum of the first element of $\bm{wx}$ (i.e., A1M1 + A2M2 + A3M3 + A4M4), and the first element in $[\bm{u}_1]_{\mathcal{C}}$ is a partial sum of the $2$nd element of $\bm{wx}$ (i.e., B1M1 + B2M2 + B3M3 + B4M4). Then, the server conducts rotations on each $[\bm{u}_i]_{\mathcal{C}}$, with totally $(n_o-1)$ Perm operations excluding the trivial rotation by zero, to make the first entry of $[\bm{u}_i]_{\mathcal{C}}$ to be a partial sum of the first element of $\bm{wx}$. Next, the server adds all of the rotated $[\bm{u}_i]_{\mathcal{C}}$ to obtain a single ciphertext whose entries are repeatedly a partial sum of the elements of $\bm{wx}$. For example, in  step (c) of Figure \ref{weight_sum_diagram3},  $[\bm{u}_1]_{\mathcal{C}}$ is rotated by one position, and then added with $[\bm{u}_0]_{\mathcal{C}}$ to get one ciphertext, whose entries are the partial sum of the first and second elements of $\bm{wx}$.

Till now, the natural next step is to conduct $\log_2\frac{n_i}{n_o}$ RaS iterations to get a final ciphertext whose first $n_o$ entries are the $n_o$ elements of $\bm{wx}$, i.e., the approach used by the hybrid method~\cite{liu2017oblivious,juvekar2018gazelle}. With GALA, we propose to eliminate the $\log_2\frac{n_i}{n_o}$ time-consuming RaS iterations by integrating it with the generation of shares for the GC-based nonlinear computing.

As introduced in the hybrid method~\cite{liu2017oblivious,juvekar2018gazelle}, in order to do the GC based nonlinear computing, the encrypted linear output is shared as follows: (1) the server generates a random vector; (2) the server subtracts the random vector from the ciphertext (the encrypted linear output); (3) the subtracted ciphertext is sent to the client, which subsequently decrypts it and obtains its share.

Here we let the server encode a similar random vector and subtract it from the ciphertext obtained in  step (c) of Figure~\ref{weight_sum_diagram3}. The subtracted ciphertext is sent to the client, which decrypts ciphertext, and then applies $\log_2{\frac{n_i}{n_o}}$  RaS iterations on the plaintext, as illustrated in step (d) of Figure~\ref{weight_sum_diagram3}. Similarly, the server gets its share by $\log_2{\frac{n_i}{n_o}}$ plaintext RaS iterations on its encoded random vector. Hence, in GALA, the server replaces the ciphertext RaS operations by much faster plaintext RaS operations. This significantly improves the computation efficiency.

Furthermore, in order to make use of all slots in a ciphertext, the client packs $\frac{n}{n_i}$ input $\bm{x}$ to form a packed vector $[\bm{x}_{\textmd{pack}}]_{\mathcal{C}}$. Then the server multiplies $\frac{n}{n_i}$ encoded weight vectors with $[\bm{x}_{\textmd{pack}}]_{\mathcal{C}}$ by one ScMult operation. As a result, the server obtains $\frac{n_in_o}{n}$ multiplied ciphertext, which are respectively rotated to enable the elementwise sum, finally producing a single ciphertext that has $\frac{n}{n_o}$ to-be-accumulated blocks.
Without any further HE RaS iterations, the server then starts to encode the random vector for the share generation. The only extra computation is the plaintext RaS iteration(s) at both the client and server, which is  much faster compared to the ones in HE domain.

As a result, GALA needs $\frac{n_in_o}{n}$ ScMult operations, $(\frac{n_in_o}{n}-1)$ Perm operations, and $(\frac{n_in_o}{n}-1)$  Add operations. It yields one output ciphertext, and makes efficient  utilization of ciphertext slots. Table \ref{complexity_matrix_vector} compares the complexity among the naive method, the hybrid method (i.e., GAZELLE) and the proposed row-encoding-share-RaS matrix-vector multiplication (GALA). We can see that the proposed method completely eliminates the HstPerm operations and significantly reduces the Perm operations.

\begin{table}[!h]
\scriptsize
\renewcommand\arraystretch{1.5}
\centering
\caption{Complexity comparison of three methods.}
\begin{tabular}{ccccc}
  \hline
  \hline
  Method & \# Perm & \# HstPerm  & \# ScMult & \# Add\\
  \hline
  \hline
  Naive & $n_o\log_2{n_i}$ & $0$ & $n_o$ & $n_o\log_2{n_i}$ \\
  \hline
  GAZELLE & $\log_2\frac{n}{n_o}$ & $\frac{n_in_o}{n}-1$ & $\frac{n_in_o}{n}$ & $\log_2\frac{n}{n_o}+\frac{n_in_o}{n}-1$ \\
  \hline
  GALA & $\frac{n_in_o}{n}-1$ & $0$ & $\frac{n_in_o}{n}$ & $\frac{n_in_o}{n}-1$ \\
  \hline
  \hline
\end{tabular}
\label{complexity_matrix_vector}
\end{table}

%% file: system_description_SecB.tex
\subsection{Kernel Grouping Based Convolution}\label{system:conv}
In this subsection, we introduce GALA's optimization for convolution. Similar to the discussion on the matrix-vector multiplication, we first begin with the basic convolution for the Single Input Single Output (SISO), then go through the state-of-the-art scheme for the Multiple Input Multiple Output (MIMO) (i.e., the GAZELLE framework~\cite{juvekar2018gazelle}). Finally we elaborate GALA's  first-Add-second-Perm ({\em kernel grouping}) scheme that achieves more efficient convolution computation.
We assume the server has $c_o$ plaintext kernels with a size of $k_w\times{k_h}\times{c_i}$ and the client sends to the server the encrypted data in the size of $u_w\times{u_h}$ with $c_i$ channels. The server needs to homomorphically convolve the encrypted data from the client with its plaintext kernels to produce the encrypted output.

\begin{figure}[!tbp]
\centering
\includegraphics[trim= {0.2cm 0cm 0cm 0cm}, clip, scale=0.30]{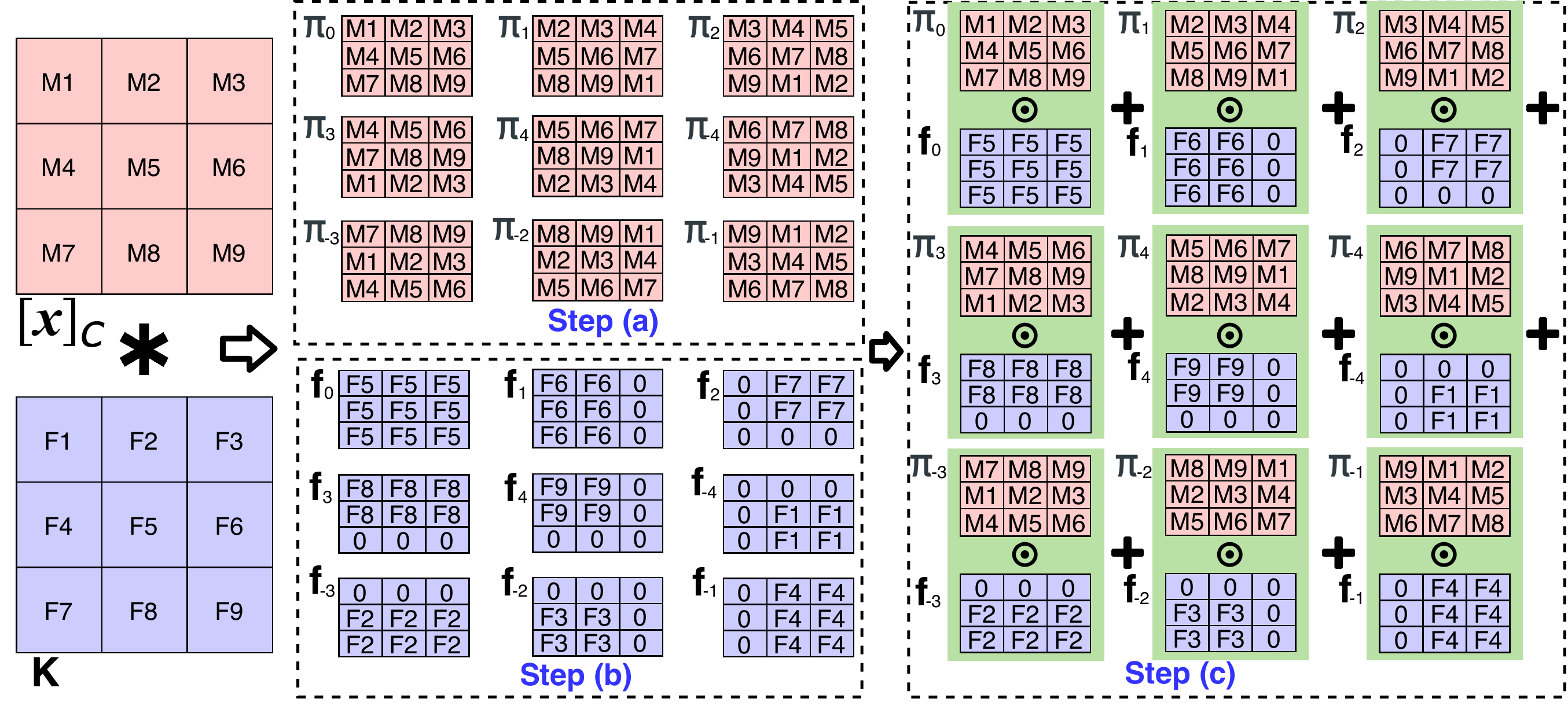}
\caption{SISO convolution.}
\label{convolution_diagram1}
\end{figure}

\noindent\textbf{1) Basic SISO convolution:}
SISO is a special case of MIMO where $c_i=c_o=1$. In this case, the encrypted data from the client has a size of $u_w\times{u_h}$ with one channel (i.e., a 2D image) and there is only one kernel with size $k_w\times{k_h}$ (i.e., a 2D filter) at the server. The SISO convolution is illustrated by an example in Figure~\ref{convolution_diagram1} where $[\bm{x}]_{\mathcal{C}}$ is the encrypted data from the client and K is the plaintext kernel at the server. The process of convolution can be visualized as placing the kernel K at different locations of the input data $[\bm{x}]_{\mathcal{C}}$. At each location, a sum of an element-wise product  between the kernel and corresponding data values within the kernel window is computed. For example, in Figure~\ref{convolution_diagram1}, the first value of the convolution between $[\bm{x}]_{\mathcal{C}}$ and kernel K is (M1F5 + M2F6 + M4F8 + M5F9). It is obtained by first placing the center of K, i.e., F5, at M1 and then calculating the element-wise product between K and the part of $[\bm{x}]_{\mathcal{C}}$ that is within K's kernel window (i.e., M1, M2, M4 and M5). The final result is the sum of the element-wise product. The rest of convolution values are calculated similarly by placing F5 at M2 to M9.

We now elaborate the convolution by an example when F5 is placed at M5 (i.e., the central element of $[\bm{x}]_{\mathcal{C}}$). In this example, the kernel size is $k_wk_h=9$. The convolution is derived by summing the element-wise product between the 9 values in K and the corresponding 9 values around M5. This can be achieved by rotating $[\bm{x}]_{\mathcal{C}}$  in a raster scan fashion~\cite{juvekar2018gazelle}. Specifically, $[\bm{x}]_{\mathcal{C}}$ is converted to a vector by concatenating all rows. Then, it is rotated by $(k_wk_h-1)$ rounds, with half of them in the forward direction and the other half in the backward direction. We denote $\pi_j$ as the rotation by $j$ positions, where a positive sign of $j$ indicates the forward direction and negative the backward direction, as shown in step (a) of Figure~\ref{convolution_diagram1}.

The convolution is obtained by (1) forming the kernel coefficients according to the partial sum at the corresponding location as shown in step (b) of Figure~\ref{convolution_diagram1}, (2) scaling the 9 rotated $\pi_j$ with the corresponding kernel coefficients, and (3) summing up all scaled $\pi_j$ (see  step (c)).

The rotation for $[\bm{x}]_{\mathcal{C}}$ is completed by HstPerm\footnote{With a common DecPerm operation.}. The scaling is done by ScMult and the summation is achieved by Add. Therefore, the SISO convolution requires a total of $(k_wk_h-1)$ HstPerm operations (excluding the trivial rotation by zero), $k_wk_h$ ScMult operations and $(k_wk_h-1)$ Add operations. The output is one ciphertext\footnote{We assume the input size $u_w{u_h}$ is smaller that the ciphertext size $n$.} which contains the convolution result.

\begin{figure}[!tbp]
\centering
\includegraphics[scale=0.26]{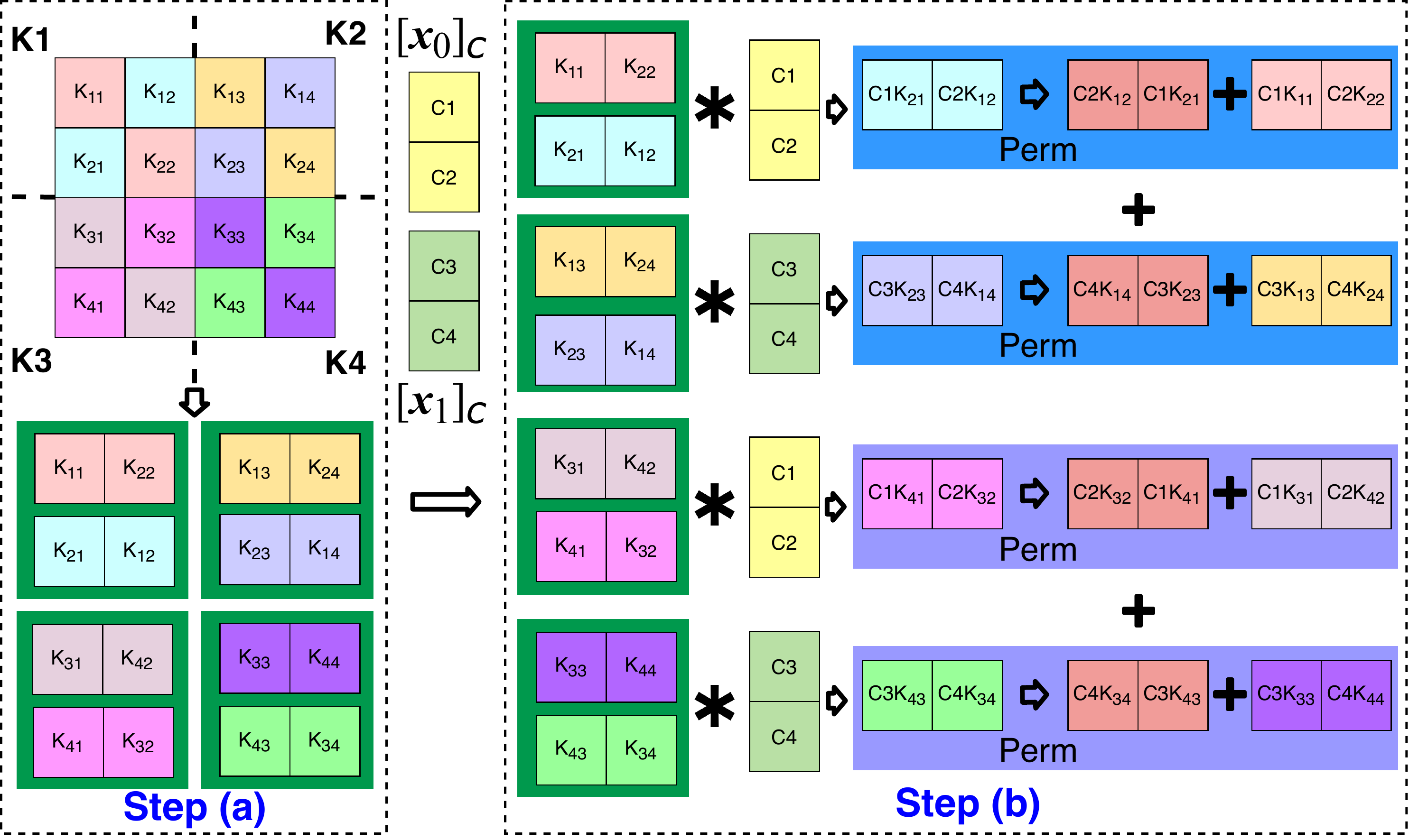}
\caption{MIMO convolution.}
\label{convolution_diagram2}
\end{figure}

\noindent\textbf{2) Output Rotation based MIMO convolution (GAZELLE):}
We now consider the more general case, i.e., MIMO, where $c_i$ or $c_o$ is not one. The naive approach is to directly apply SISO convolution by first encrypting the $c_i$ input channels into $c_i$ ciphertext, $\{[\bm{x}_i]_{\mathcal{C}}\}$. Each of the $c_o$ kernels includes $c_i$ filters. Each $[\bm{x}_i]_{\mathcal{C}}$ is convolved with one of the $c_i$ filters by SISO and the final convolution is obtained by summing up all of the $c_i$ SISO convolutions. As a result, the naive approach  requires $c_i(k_wk_h-1)$ HstPerm operations (for $c_i$ input channels), $c_ic_ok_wk_h$ ScMult operations and $c_o(c_ik_wk_h-1)$ Add operations. There are  $c_o$ output ciphertext.

Given the number of slots $n$ in a ciphertext is usually larger than the channel size $u_w{u_h}$,
the ciphertext utilization (i.e., the meaningful slots that output desired results) in the $c_o$ output ciphertext is low.

In order to improve the ciphertext utilization and computation efficiency for MIMO convolution,
the state-of-the-art method (i.e., the output rotation~\cite{juvekar2018gazelle}) first packs $c_n$  channels of input data into one ciphertext, which results in $\frac{c_i}{c_n}$ input ciphertext (see Figure~\ref{convolution_diagram2} where the four input channels form two ciphertext, each of which includes two channels). Meanwhile, the $c_o$ kernels are viewed as a $c_o\times{c_i}$ kernel block and each row of the block includes $c_i$ 2D filters for one kernel. Then the MIMO convolution is viewed as a matrix-vector multiplication where the element-wise multiplication is replaced by convolution. As each ciphertext holds $c_n$ channels, the kernel block is divided into $\frac{c_oc_i}{c_n^2}$ blocks (see step (a) in Figure~\ref{convolution_diagram2}, where the kernel block is divided into K1 to K4).

Next, each divided block is diagonally encoded into $c_n$ vectors such that the first filters in all vectors are in the first column of the kernel block (see the four groups of vectors in step (a) of Figure~\ref{convolution_diagram2}). In this way, each input ciphertext can directly convolve with the vectors in each divided block by SISO, and the convolution for each divided block is obtained by rotating the $c_n$ convolved vectors to the same kernel order as the diagonal one and summing them up (see step (b)).

Finally, the convolution for $c_n$ kernels is calculated by adding the convolution of $\frac{c_i}{c_n}$ blocks associated with the same kernels as illustrated in step (b) of Figure~\ref{convolution_diagram2}.

Clearly, there are $\frac{c_o}{c_n}$ output ciphertext, as expected. For each of the $\frac{c_oc_i}{c_n^2}$ blocks, there are total $c_n$ SISO-like convolutions, requiring $c_nk_wk_h$ ScMult operations, $(c_n-1)$ Perm operations and $(c_nk_wk_h-1)$ Add operations. Next, there are $\frac{c_i}{c_n}$ block convolutions which are associated with the same kernel order. Thus they are added up to obtain the final convolution result. Meanwhile, the rotation group for each input ciphertext is reused to convolve with different kernel blocks. Thus there are total $\frac{c_i(k_wk_h-1)}{c_n}$ HstPerm operations with $\frac{c_i}{c_n}$ common DecPerm operations.
In all, the MIMO convolution needs a total of $\frac{c_ic_o}{c_n^2}(c_n-1)$ Perm,
$\frac{c_i}{c_n}(k_wk_h-1)$ HstPerm, $k_wk_h\frac{c_ic_o}{c_n}$ ScMult and $\frac{c_o}{c_n}(c_ik_wk_h-1)$ Add operations.

\begin{figure}[!tbp]
\centering
\includegraphics[scale=0.26]{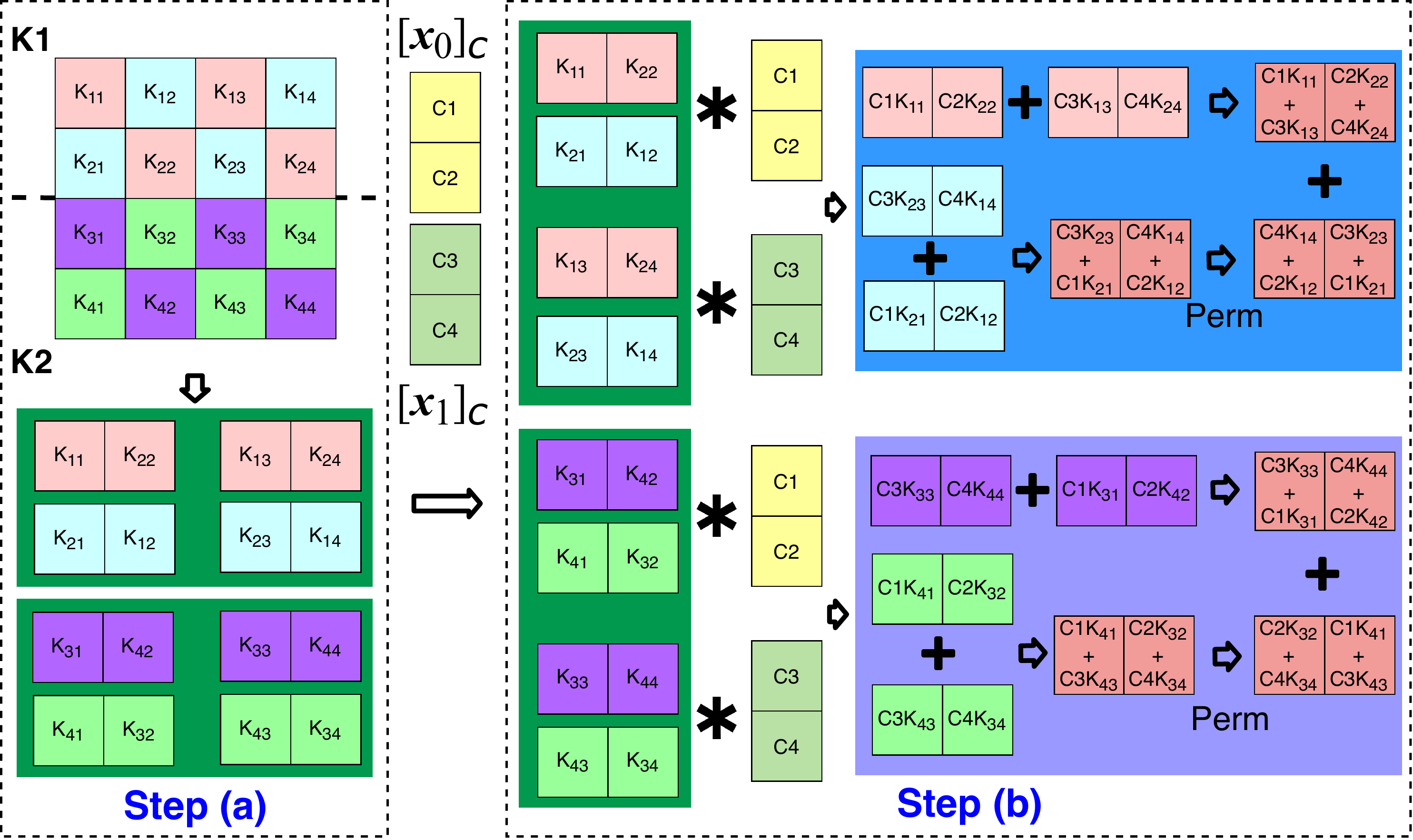}
\caption{Kernel grouping based MIMO convolution.}
\label{convolution_diagram3}
\end{figure}

\noindent\textbf{3) Kernel Grouping Based MIMO convolution (GALA):}
One key observation on the above MIMO convolution is that,
each of the $\frac{c_oc_i}{c_n^2}$ blocks needs $(c_n-1)$ expensive Perm operations in order to get the convolution for that block. However, we actually do not need to get the convolution for each block. As our goal is to get the convolution for each kernel, the blocks that are associated with the same kernel are combined in our proposed first-Add-second-Perm approach ({\em kernel grouping}) to reduce the Perm cost. Specifically, in step (a) of Figure~\ref{convolution_diagram3}, the whole kernel block is divided into two blocks K1 and K2 such that each block is the combination of $\frac{c_i}{c_n}$ $c_n$-by-$c_n$ divided blocks, which  correspond  to the same kernels (i.e., the first and second kernel in K1 and the third and fourth kernel in K2).

For each newly formed block, all of the vectors are first convolved with the corresponding input ciphertext by SISO-like convolution. Then the convolved vectors that are associated with the same kernel order are first added together (see the addition of convolved vectors before rotation in step (b) of Figure~\ref{convolution_diagram3}). Finally, these added vectors are rotated to the same kernel order and summed up to obtain the convolution result (see the rotation and final addition for each block in step (b) of Figure~~\ref{convolution_diagram3}).

This kernel grouping calculation results in $(c_n-1)$ Perm operations for each of $\frac{c_o}{c_n}$ newly formed blocks, which
reduces the Perm complexity by a factor of $\frac{c_i}{c_n}$ compared with GAZELLE's MIMO convolution. This reduction is nontrivial especially for the state-of-the-art neural networks such as ResNets \cite{he2016deep}, where $\frac{c_i}{c_n}$ can be 256. This is because these neural networks contain a large number of large-size feature maps in order to capture the complex input features~\cite{simonyan2014very, krizhevsky2012imagenet, he2016deep}.

Similar to the output rotation based MIMO convolution discussed above, there are $\frac{c_o}{c_n}$ output ciphertext in the proposed scheme. For each of the $\frac{c_o}{c_n}$ newly formed blocks, there are $c_i$ SISO-like convolutions. Then for each of the $c_n$ kernel orders, there are $\frac{c_i}{c_n}$ convolutions to be summed up, which results in $c_n$ added convolutions. These added convolutions are further rotated to the same kernel order and summed up to get the final convolution.
Therefore, the proposed MIMO convolution requires a total of $\frac{c_o}{c_n}(c_n-1)$ Perm,
$\frac{c_i}{c_n}(k_wk_h-1)$ HstPerm, $k_wk_h\frac{c_ic_o}{c_n}$ ScMult, and $\frac{c_o}{c_n}(c_ik_wk_h-1)$ Add operations.

Table~\ref{complexity_conv} compares the overall complexity for convolution computations. GALA's kernel grouping approach reduces the expensive Perm operations by a factor of $\frac{c_i}{c_o}$ without increasing other operations compared with the output rotation based MIMO convolution (i.e., the GAZELLE framework). The reduction in Perm operations leads to a significant speedup. Specifically, GALA shows about 14$\times$ speedup compared with GAZELLE in the convolution between  input data with a size of 16$\times$16 with 2048 channels, and 512 kernels with a size of  1$\times$1$\textit{@}$2048 on a commodity machine (see detailed benchmarks in Sec. \ref{evaluation}).

\begin{table}[!h]
{\scriptsize
\renewcommand\arraystretch{1.5}
\centering
\caption{Complexity comparison of convolution.}
\begin{tabular}{ccccc}
  \hline
  \hline
  Method & \# Perm & \# HstPerm$^{\sharp}$  & \# ScMult & \# Add\\
  \hline
  \hline
  GAZELLE & $\frac{c_ic_o(c_n-1)}{c_n^2}$ & $\frac{c_i(k_wk_h-1)}{c_n}$ & $\frac{c_ic_ok_wk_h}{c_n}$ & $\frac{c_o(c_ik_wk_h-1)}{c_n}$ \\
  \hline
  GALA & $\frac{c_o(c_n-1)}{c_n}$ & $\frac{c_i(k_wk_h-1)}{c_n}$ & $\frac{c_ic_ok_wk_h}{c_n}$ & $\frac{c_o(c_ik_wk_h-1)}{c_n}$ \\
  \hline
  \hline
  \multicolumn{5}{l}{$^{\sharp}$Rotations of the input with $\frac{c_i}{c_n}$ common DecPerm operations.}
\end{tabular}
\label{complexity_conv}
}
\end{table}

%% file: system_description_SecC.tex
\subsection{Noise Management}\label{noise_manage}
The packed HE (e.g., the BFV scheme) introduces noise in the ciphertext which theoretically hides the original message \cite{juvekar2018gazelle, brakerski2012fully}. However, the noise management is critical to the correct decryption of ciphertext after a series of HE operations. We will show that GALA has better noise management compared with GAZELLE.

Based on the computation complexity of matrix-vector multiplication and convolution, along with the noise change for HE operations as described in Sec.~\ref{priliminary:crypto}, Table~\ref{noise_growth_complexity} shows the noise growth of different schemes. As for the matrix-vector multiplication, GALA has a lower noise growth while keeping the number of output ciphertext as small as one\footnote{Note that the noise in Table~\ref{noise_growth_complexity} is calculated by assuming $(\frac{n_in_o}{n}-1)\geq0$. The noise of GALA is still lower than that of GAZELLE when $(\frac{n_in_o}{n}-1)<0$ as it means one ciphertext can hold data with size $n_o\times{n_i}$, which only involves one ScMult operation in GALA,  and GAZELLE needs to subsequently conduct a series of RaS operations.}.
As for the convolution computation, GALA reduces the noise term associated with rotation by a factor of $\frac{c_i}{c_n}$ compared to GAZELLE. This is nontrivial especially for  state-of-the-art neural networks such as ResNets~\cite{he2016deep}, where $\frac{c_i}{c_n}$ can be 256. The number of output ciphertext is also maintained as small as $\frac{c_o}{c_n}$. Overall, GALA features a lower noise growth and lower computation complexity compared with GAZELLE.

\begin{table}[!htbp]
\footnotesize
\centering
\caption{Comparison of noise management.}
\begin{tabular}{c|c|c}
\hline
\hline
\multicolumn{3}{c}{Matrix-vector Multiplication}\\
\hline
Method & Noise after computation & \# Cipher\\
\hline
Naive & $n_i\eta_0\eta_{mult}+(n_i-1)\eta_{rot}$ & $n_o$\\
GAZELLE & $n_i\eta_0\eta_{mult}+[\frac{n_in_o-n}{n_o}\eta_{mult}+\frac{n-n_o}{n_o}]\eta_{rot}$ & 1\\
GALA & $\frac{n_in_o}{n}\eta_0\eta_{mult}+(\frac{n_in_o}{n}-1)\eta_{rot}$ & 1\\
\hline
\hline
\multicolumn{3}{c}{Convolution Computation}\\
\hline
Method & Noise after computation & \# Cipher\\
\hline
GAZELLE & $c_i\eta_{\Delta}+\frac{c_i}{c_n}(c_n-1)\eta_{rot}$ & $\frac{c_o}{c_n}$\\
GALA & $c_i\eta_{\Delta}+(c_n-1)\eta_{rot}$ & $\frac{c_o}{c_n}$\\
\hline
\hline
\multicolumn{3}{l}{$\eta_{\Delta}=k_wk_h\eta_{mult}\eta_0+(k_wk_h-1)\eta_{rot}\eta_{mult}$}
\end{tabular}\label{noise_growth_complexity}
\end{table}

%% file: system_description_SecD.tex
\subsection{System Security}

GALA is based on the same security framework as GAZELLE~\cite{juvekar2018gazelle}.
The security of linear computation in GALA is fully protected by the security of HE (e.g., the BFV scheme~\cite{brakerski2012fully, fan2012somewhat}). The nonlinear computation (which is not the focus of this paper) is protected by Garbled Circuits (GC)~\cite{yao1986generate} or its alternatives.
{\color{black}The security of GC-based nonlinear computation has been proven in TASTY~\cite{henecka2010tasty} and MP2ML~\cite{boemer2020mp2ml}.}

%% file: evaluation.tex
\section{Evaluation}\label{evaluation}
{\color{black}
We conduct the experiments in both LAN and WAN settings. The LAN setting is implemented on a Gigabit Ethernet in our lab between two workstations as the client and server, respectively. Both machines run Ubuntu, and have an Intel i7-8700 3.2GHz CPU with 12 threads and 16 GB RAM. The WAN setting is based on a connection between a local PC and an Amazon AWS server with an average bandwidth  200Mbps and round-trip time around 13ms. We have downloaded the codes released by GAZELLE\footnote{Available at https://github.com/chiraag/gazelle\_mpc}, DELPHI\footnote{\color{black}Available at https://github.com/mc2-project/delphi} and CrypTFlow2\footnote{\color{black}Available at https://github.com/mpc-msri/EzPC/tree/master/SCI}, and run all experiments on the same hardware devices and network settings. We conduct a series of experiments under various neural network architectures. In each experiment, we first run the baseline algorithm (i.e., GAZELLE, DELPHI or CrypTFlow2) to obtain the baseline total runtime (including online runtime and offline runtime), and then replace the linear computation of the baseline algorithm by GALA to get a new total runtime, which is then used to compute the speedup.

While the codes for GAZELLE, DELPHI and CrypTFlow2 are implemented in different ways (for example, GAZELLE is based on its crypto platform while DELPHI and CrypTFlow2 are based on the Microsoft SEAL library), we focus on the speedup of GALA on top of each of them.}
We also set the cryptographic parameters in line with GAZELLE: 1) Parameters for both HE and GC schemes are selected for a 128-bit security level. 2) A plaintext modulus $p$ of 20 bits is enough to store all the intermediate values in the network computation. 3) The ciphertext modulus $q$ is chosen to be a 60-bit pseudo-Mersenne prime that is slightly smaller
than the native machine word on a 64-bit machine to enable
lazy modular reductions. 4) The selection of the number of slots is the smallest power of two that allows for a 128-bit security which in our case is $n = 2048$. We refer readers to~\cite{juvekar2018gazelle} for more details about the parameter selection.

\subsection{Microbenchmarks}
In this section, we benchmark and compare the runtime of GALA's linear optimization (i.e., matrix-vector multiplication and convolution computation) with state-of-the-art approaches. We claim the same communication cost and inference accuracy with GAZELLE and achieve improved computation efficiency.

\noindent\textbf{1) Matrix-Vector Multiplication:}
Table~\ref{mv_performance} compares the computation complexity of GALA's matrix-vector optimization with GAZELLE and two other optimization schemes (i.e., a diagonal method (Diagonal)~\cite{halevi2014algorithms} and an extended method (Extended)~\cite{chen2019efficient}). We can see that GALA largely reduces the expensive Perm operation to zero in our cases (including the HstPerm) while GAZELLE needs up to 11 Perm and Extended~\cite{chen2019efficient} needs up to 520 Perm (including HstPerm). On the other hand, GALA also maintains a light overhead for HE multiplication/addition, i.e., only one multiplication, compared with other three optimizations, e.g., Diagonal~\cite{halevi2014algorithms} and Extended~\cite{chen2019efficient} involve up to 2048 multiplication/addtion.

{\color{black}The runtime results for matrix-vector multiplication are summarized in Table~\ref{mv_performance_time}, which includes the original runtime of GAZELLE, DELPHI and CrypTFlow2, and the speedup of GALA on top of each.} We take the share-RaS calculation cost (see the plaintext computing for final share at the client in step (d) of Figure~\ref{weight_sum_diagram3}) as part of the runtime cost of GALA for fair comparison. Meanwhile, as multiple copies are packed in one ciphertext, the HstPerm operation includes a common DecPerm to enable hoist optimization for rotation (see the details in~\cite{juvekar2018gazelle}). As can be seen from Table~\ref{mv_performance_time}, GALA's optimization gains a large speedup due to the row-encoding-share-RaS module, which reduces the costly Perm, Mult, and Add operations for a series of RaS calculation. Specifically, GALA achieves the speedup of 1795$\times$, 208$\times$ and 57$\times$ over the Diagonal~\cite{halevi2014algorithms} under different matrix dimensions in the LAN setting. This benefit stems from the fact that the computation complexity of the Diagonal is related to the input dimension $n_i$, which is always large in the state-of-the-art neural networks such as AlexNet~\cite{krizhevsky2012imagenet}, VGG~\cite{simonyan2014very} and ResNet~\cite{he2016deep}. For a similar reason, GALA significantly outperforms the Extended method~\cite{chen2019efficient}.

Meanwhile, GALA has a speedup of 59$\times$, 13$\times$ and 19$\times$ over GAZELLE under different matrix dimensions in the LAN setting. This computation gain comes from the HstPerm-free scheme
(i.e., row-encoding) and elimination of RaS computation (i.e., share-RaS scheme) compared to GAZELLE, which is particularly effective for large $\frac{n_i}{n_o}$ ratio and large ciphertext slots (see the superior performance for the neural network with a dimension of $1\times2048$). These features suit well to current convolutional neural networks which have tens of thousands of values to be fed into the fully connected layers~\cite{simonyan2014very, he2016deep}.

{\color{black}Compared with DELPHI and CrypTFlow2, GALA achieves a speedup for weight matrix multiplication up to 700$\times$ in the LAN setting.
This is largely due to GALA's deep optimization for HE computation. We also notice that GALA's speedup slows down in WAN which is due to the communication rounds needed for conversions between HE and GC. Therefore it leads to significant round time in total
compared with the light HE computation overhead. For example, the round-trip time is around 13 milliseconds while the GALA's optimized HE cost is within one millisecond.
}

\begin{table}[!tbp]
\footnotesize
\centering
\caption{Computation complexity of matrix-vector multiplication.}
\begin{tabular}{c|c|c|c|c}
\hline
\hline
\multicolumn{5}{c}{Dimension ($n_o\times{n_i}$): 1$\times$2048}\\
\hline
Metric & Diagonal\cite{halevi2014algorithms} & GAZELLE & Extended\cite{chen2019efficient} & GALA\\
\hline
\# Perm & 0 & 11 & 0& 0\\
\# HstPerm$^{\natural}$ & 2047 & 0 & 2047& 0\\
\# ScMult & 2048 & 1 & 2048& 1\\
\# Add & 2047 & 11 & 2047& 0\\
\hline
\hline
\multicolumn{5}{c}{Dimension ($n_o\times{n_i}$): 2$\times$1024}\\
\hline
Metric & Diagonal\cite{halevi2014algorithms} & GAZELLE & Extended\cite{chen2019efficient} & GALA\\
\hline
\# Perm & 0 & 10 & 9& 0\\
\# HstPerm$^{\natural}$ & 1023 & 0 & 511& 0\\
\# ScMult & 1024 & 1 & 512& 1\\
\# Add & 1023 & 10 & 520& 0\\
\hline
\hline
\multicolumn{5}{c}{Dimension ($n_o\times{n_i}$): 16$\times$128}\\
\hline
Metric & Diagonal\cite{halevi2014algorithms} & GAZELLE & Extended\cite{chen2019efficient} & GALA\\
\hline
\# Perm & 0 & 7 & 4& 0\\
\# HstPerm$^{\natural}$ & 127 & 0 & 7& 0\\
\# ScMult & 128 & 1 & 8& 1\\
\# Add & 127 & 7 & 11& 0\\
\hline
\hline
\multicolumn{5}{l}{$^{\natural}$Rotations of the input with a common DecPerm}
\end{tabular}\label{mv_performance}
\end{table}

\begin{table}[!tbp]
\footnotesize
\centering
\caption{Runtime cost of matrix-vector multiplication.}
\begin{tabular}{c|c|c|c|c|c}
\hline
\hline
\multicolumn{6}{c}{Dimension ($n_o\times{n_i}$): 1$\times$2048}\\
\hline
\multirow{2}{*}{Approach} & \color{black}Comm. & \multicolumn{2}{c|}{LAN (ms)} & \multicolumn{2}{c}{\color{black}WAN (ms)}\\
 & \color{black}(MB) & Time & \textbf{Speedup} & \color{black}Time & \color{black}\textbf{Speedup} \\
\hline
Diagonal\cite{halevi2014algorithms} & \color{black}0.03 &57 & \textbf{1795$\times$}&\color{black}75 & \color{black}\textbf{4$\times$}\\
Extended\cite{chen2019efficient} & \color{black}0.03 &57.5 & \textbf{1796$\times$}& \color{black}77 &\color{black}\textbf{4$\times$} \\
GAZELLE\cite{juvekar2018gazelle} & \color{black}0.03 & 1.9& \textbf{59$\times$}& \color{black}19.3& \color{black}{1$\times$}\\
\color{black}DELPHI\cite{mishra2020delphi} & \color{black}0.14&\color{black}28 &\color{black}\textbf{700$\times$} & \color{black}59.5& \color{black}\textbf{3.2$\times$}\\
\color{black}CrypTFlow2\cite{rathee2020cryptflow2} & \color{black}0.13 & \color{black}28&\color{black}\textbf{700$\times$} & \color{black}46.2 & \color{black}\textbf{2.5$\times$}\\
\hline
\hline
\multicolumn{6}{c}{Dimension ($n_o\times{n_i}$): 2$\times$1024}\\
\hline
Diagonal\cite{halevi2014algorithms} & \color{black}0.03 &28 & \textbf{208$\times$}&\color{black}47 & \color{black}\textbf{2.5$\times$}\\
Extended\cite{chen2019efficient} & \color{black}0.03 &16 & \textbf{116$\times$}& \color{black}36 &\color{black}\textbf{1.9$\times$} \\
GAZELLE\cite{juvekar2018gazelle} & \color{black}0.03 & 1.8& \textbf{13$\times$}& \color{black}19& \color{black}{1$\times$}\\
\color{black}DELPHI\cite{mishra2020delphi} & \color{black}0.13&\color{black}26.5 &\color{black}\textbf{176$\times$} & \color{black}57.8& \color{black}\textbf{3.1$\times$}\\
\color{black}CrypTFlow2\cite{rathee2020cryptflow2} & \color{black}0.13 & \color{black}26.5&\color{black}\textbf{176$\times$} & \color{black}44.7 & \color{black}\textbf{2.4$\times$}\\
\hline
\hline
\multicolumn{6}{c}{Dimension ($n_o\times{n_i}$): 16$\times$128}\\
\hline
Diagonal\cite{halevi2014algorithms} & \color{black}0.03 &3.7 & \textbf{57$\times$}&\color{black}21 & \color{black}{1$\times$}\\
Extended\cite{chen2019efficient} & \color{black}0.03 &1 & \textbf{16$\times$}& \color{black}20.4 &\color{black}{1$\times$} \\
GAZELLE\cite{juvekar2018gazelle} & \color{black}0.03 & 1.2& \textbf{19$\times$}& \color{black}21& \color{black}{1$\times$}\\
\color{black}DELPHI\cite{mishra2020delphi} & \color{black}0.13&\color{black}20.5 &\color{black}\textbf{292$\times$} & \color{black}51.7& \color{black}\textbf{2.8$\times$}\\
\color{black}CrypTFlow2\cite{rathee2020cryptflow2} & \color{black}0.13 & \color{black}20.5&\color{black}\textbf{292$\times$} & \color{black}38.7 &\color{black} \textbf{2.1$\times$}\\
\hline
\hline
\end{tabular}\label{mv_performance_time}
\end{table}

\noindent\textbf{2) Convolution Computation:}
{\color{black}
We benchmark and compare the computation complexity and runtime of GALA with GAZELLE, DELPHI and CrypTFlow2 for convolution calculation. The details are illustrated in Table~\ref{cov_performance} and~\ref{conv_performance_time}.}
As for the computation complexity,
we compare GALA with GAZELLE whose privacy-preserved convolution calculation over HE is one of the most optimized methods in current literature.
While introducing no extra HE multiplication/addition, GALA reduces the most expensive Perm, i.e., DecPerm and HstPerm, by up to 59$\times$ for input size of 16$\times$16\textit{@}2048 with kernel size of 1$\times$1\textit{@}512. This block  with large channels and small kernel size is featured in state-of-the-art neural networks such as ResNets~\cite{he2016deep}, which makes GALA suitable to boost the modern networks.

As for runtime comparison shown in Table~\ref{conv_performance_time}, GALA demonstrates 9$\times$, 14$\times$ and 2.6$\times$ speedup over GAZELLE with different input and kernel dimensions in LAN setting. As analyzed in Sec.~\ref{system:conv}, due to the fundamental complexity reduction by GALA's kernel grouping approach,
GALA reduces the expensive Perm operation by a factor of $\frac{c_i}{c_n}$. As we mention above, the large speedup is achieved under large input channels and small kernel size, the proposed approach fits very well with the state-of-the-art networks such as ResNets~\cite{he2016deep}, where the feature maps are always with large channels (which results in large $c_i$ while $c_n$ is fixed) and small kernels (that are usually 1$\times$1, 3$\times$3 and 5$\times$5 at most, which benefit small HE multiplication/addition). {\color{black} Meanwhile, the speedup over DELPHI and CrypTFlow2 is up to 7.4$\times$ in the LAN setting. On the other hand, the speedup of GALA in the   WAN setting is also decent, up to 8.7$\times$, 6.3$\times$ and 6.5$\times$ for GAZELLE, DELPHI and CrypTFlow2, respectively. This is because the computation cost of convolution increases accordingly with regard to  the communication cost, compared with the case of  matrix-vector multiplication.
}

\begin{table}[!tbp]
\footnotesize
\centering
\caption{Computation complexity of convolution.}
\begin{tabular}{c|c|c|c|c}
\hline
\hline
Input $^{\dag}$ & Kernel $^{\ddag}$ & Metric & GAZELLE\cite{juvekar2018gazelle}& GALA\\
\hline
\multirow{4}{*}{16$\times$16$\textit{@}$128} & \multirow{4}{*}{1$\times$1$\textit{@}$128} & \# DecPerm & 1792& 112\\
 & & \# HstPerm & 1792& 112\\
 & & \# ScMult & 2048& 2048\\
 & & \# Add & 2032& 2032\\
\hline
\hline
\multirow{4}{*}{16$\times$16$\textit{@}$2048} & \multirow{4}{*}{1$\times$1$\textit{@}$512} & \# DecPerm & 114944 & 2048\\
 & & \# HstPerm & 114688 & 1792\\
 & & \# ScMult & 131072 & 131072\\
 & & \# Add & 130944 & 130944\\
\hline
\hline
\multirow{4}{*}{16$\times$16$\textit{@}$128} & \multirow{4}{*}{3$\times$3$\textit{@}$128} & \# DecPerm & 1808& 128\\
 & & \# HstPerm & 1920& 240\\
 & & \# ScMult & 18432& 18432\\
 & & \# Add & 18416& 18416\\
\hline
\hline
\multirow{4}{*}{\color{black}16$\times$16$\textit{@}$2048} & \multirow{4}{*}{\color{black}5$\times$5$\textit{@}$64} & \color{black}\# DecPerm & \color{black}14592 & \color{black}312\\
 & & \color{black}\# HstPerm &\color{black} 20480 & \color{black}6200\\
 & & \color{black}\# ScMult & \color{black}409600& \color{black}409600\\
 & & \color{black}\# Add & \color{black}409592& \color{black}409592\\
\hline
\hline
\multicolumn{5}{l}{$^{\dag}$Dim. is in the form of $u_w\times{u_h}\textit{@}c_i$}\\
\multicolumn{5}{l}{$^{\ddag}$Dim. is in the form of $k_w\times{k_h}\textit{@}c_o$ with $c_i$ channels per kernel}
\end{tabular}\label{cov_performance}
\end{table}

\begin{table}[!t]
\footnotesize
\centering
\caption{Runtime cost of convolution.}
\begin{tabular}{c|c|c|c|c|c}
\hline
\hline
\multicolumn{6}{c}{Dimension (Input Dim.$^{\dag}$, Kernel Dim.$^{\ddag}$): 16$\times$16$\textit{@}$128, 1$\times$1$\textit{@}$128}\\
\hline
\multirow{2}{*}{Approach} & \color{black}Comm. & \multicolumn{2}{c|}{LAN (ms)} & \multicolumn{2}{c}{\color{black}WAN (ms)}\\
 & \color{black}(MB) & Time & \textbf{Speedup} & \color{black}Time &\color{black} \textbf{Speedup} \\
\hline
GAZELLE & \color{black}0.5 & 321& \textbf{9$\times$}& \color{black}408& \color{black}\textbf{3.2$\times$}\\
\color{black}DELPHI & \color{black}2.1&\color{black}391 &\color{black}\textbf{3.1$\times$} &\color{black} 502& \color{black}\textbf{2.3$\times$}\\
\color{black}CrypTFlow2 & \color{black}2 & \color{black}389&\color{black}\textbf{3.1$\times$} &\color{black} \color{black}482 & \color{black}\textbf{2.2$\times$}\\
\hline
\hline
\multicolumn{6}{c}{Dimension (Input Dim.$^{\dag}$, Kernel Dim.$^{\ddag}$): 16$\times$16$\textit{@}$2048 , 1$\times$1$\textit{@}$512}\\
\hline
GAZELLE & \color{black}8 & 20583.5& \textbf{14$\times$}& \color{black}21784& \color{black}\textbf{8.7$\times$}\\
\color{black}DELPHI & \color{black}31&\color{black}17939 &\color{black}\textbf{4.4$\times$} & \color{black}19205& \color{black}\textbf{3.7$\times$}\\
\color{black}CrypTFlow2 & \color{black}29 & \color{black}17928&\color{black}\textbf{4.4$\times$} & \color{black}19101 & \color{black}\textbf{3.6$\times$}\\
\hline
\hline
\multicolumn{6}{c}{Dimension (Input Dim.$^{\dag}$, Kernel Dim.$^{\ddag}$): 16$\times$16$\textit{@}$128, 3$\times$3$\textit{@}$128}\\
\hline
GAZELLE & \color{black}0.5 & 457& \textbf{2.6$\times$}& \color{black}547& \color{black}\textbf{2.1$\times$}\\
\color{black}DELPHI & \color{black}2&\color{black}2563.6 &\color{black}\textbf{5.8$\times$} & \color{black}2671& \color{black}\textbf{5$\times$}\\
\color{black}CrypTFlow2 & \color{black}1.9 & \color{black}2559&\color{black}\textbf{5.8$\times$} & \color{black}2648 & \color{black}\textbf{5$\times$}\\
\hline
\hline
\multicolumn{6}{c}{\color{black}Dimension (Input Dim.$^{\dag}$, Kernel Dim.$^{\ddag}$): 16$\times$16$\textit{@}$2048, 5$\times$5$\textit{@}$64}\\
\hline
\color{black}GAZELLE & \color{black}8 & \color{black}5875.2& \color{black}\textbf{ 1.7$\times$}& \color{black}7073& \color{black}\textbf{ 1.5$\times$}\\
\color{black}DELPHI & \color{black}31&\color{black}56499 &\color{black}\textbf{ 7.4$\times$} & \color{black}57765& \color{black}\textbf{ 6.3$\times$}\\
\color{black}CrypTFlow2 & \color{black}29 & \color{black}56409&\color{black}\textbf{ 7.4$\times$} & \color{black}57582 & \color{black}\textbf{ 6.5$\times$}\\
\hline
\hline
\multicolumn{6}{l}{$^{\dag}$Dim. is in the form of $u_w\times{u_h}\textit{@}c_i$}\\
\multicolumn{6}{l}{$^{\ddag}$Dim. is in the form of $k_w\times{k_h}\textit{@}c_o$ with $c_i$ channels per kernel}
\end{tabular}\label{conv_performance_time}
\end{table}

\subsection{Performance with Classic Networks}

\begin{table}[!t]
\footnotesize
\centering
\caption{Computation complexity of state-of-the-art neural network models.}
\begin{tabular}{c|c|c|c}
\hline
\hline
Net. Frameworks & Metric & GAZELLE\cite{juvekar2018gazelle} & GALA\\
\hline
\multirow{3}{*}{\color{black}MLP}& \color{black}\# Perm & \color{black}70 & \color{black}55\\
 &\color{black}\# ScMult & \color{black}56 & \color{black}56\\
 &\color{black}\# Add & \color{black}70 & \color{black}55\\
\hline
\hline
\multirow{5}{*}{AlexNet}& \# Perm & 40399 & 1157\\
 &\# DecPerm & 143 & 142\\
 &\# HstPerm & 1493 & 1492\\
 &\# ScMult & 481298 & 481298\\
 &\# Add & 481096 & 481089\\
\hline
\hline
\multirow{5}{*}{VGG} & \# Perm & 66055 & 2115\\
 & \# DecPerm & 161 & 160\\
 &\# HstPerm & 1283 & 1280\\
 &\# ScMult & 663556 & 663556\\
 &\# Add & 663370 & 663363\\
\hline
\hline
\multirow{5}{*}{ResNet-18}& \# Perm & 180375 & 5921\\
 & \# DecPerm & 483 & 482\\
 &\# HstPerm & 3467 & 3464\\
 &\# ScMult & 1399363 & 1399363\\
 &\# Add & 1398778 & 1398771\\
\hline
\hline
\multirow{5}{*}{ResNet-50}& \# Perm & 1464119 & 30615\\
 &\# DecPerm & 2819 & 2818\\
 &\# HstPerm & 3863 & 3848\\
 &\# ScMult & 2935408 & 2935408\\
 &\# Add & 2931734 & 2931727\\
\hline
\hline
\multirow{5}{*}{ResNet-101} & \# Perm & 2560823 & 64887\\
 & \# DecPerm & 6083 & 6082\\
 &\# HstPerm & 8215 & 8200\\
 &\# ScMult & 5302896 & 5302896\\
 &\# Add & 5294326 & 5294319\\
\hline
\hline
\multirow{5}{*}{ResNet-152}& \# Perm & 3463991 & 95127\\
 & \# DecPerm & 8963 & 8962\\
 &\# HstPerm & 12055 & 12040\\
 &\# ScMult & 7252592 & 7252592\\
 &\# Add & 7239894 & 7239887\\
\hline
\hline
\end{tabular}\label{overall_nets_num}
\end{table}

\begin{table}[!t]
\footnotesize
\centering
\caption{\color{black}Runtime cost of classic model.}
{\scriptsize
\begin{tabular}{c|c|c|c|c|c}
\hline
\hline
\multicolumn{6}{c}{\color{black}Network Model: MLP}\\
\hline
\multirow{2}{*}{\color{black}Approach} & \color{black}Comm. & \multicolumn{2}{c|}{\color{black}LAN (ms)} & \multicolumn{2}{c}{\color{black}WAN (ms)}\\
 &\color{black} (MB) & \color{black}Time &\color{black} \textbf{Speedup} & \color{black}Time & \color{black}\textbf{Speedup} \\
\hline
\color{black}SecureML & \color{black}0.21  & \color{black}31.9  & \color{black}{  \textbf{2.6}$\times$} &\color{black}79.3   & \color{black}\textbf{1.5$\times$}\\
\color{black}MiniONN & \color{black}4.4  & \color{black}14.1  & \color{black}{  1$\times$} &\color{black}227.6  & \color{black}{ {1}$\times$}\\
\color{black}GAZELLE & \color{black}0.21  & \color{black}15 &\color{black} {  1$\times$}& \color{black}84.9 &\color{black} {  1$\times$}\\
\color{black}DELPHI & \color{black}84 & \color{black}204.5 & \color{black}\textbf{  3.1$\times$} &\color{black} 3658.3 & \color{black}{ 1$\times$}\\
\color{black}CrypTFlow2 & \color{black}12.4  & \color{black}246 &\color{black}\textbf{  2.3$\times$} &  \color{black}780.6 & \color{black} {  1.2$\times$}\\
\hline
\hline
\end{tabular}\label{mlp_performance_time}
}
\end{table}

\begin{table}[!t]
\footnotesize
\centering
\caption{Runtime cost of state-of-the-art models.}
{\scriptsize
\begin{tabular}{c|c|c|c|c|c}
\hline
\hline
\multicolumn{6}{c}{Network Model: AlexNet}\\
\hline
\multirow{2}{*}{Approach} & Comm. & \multicolumn{2}{c|}{LAN (ms)} & \multicolumn{2}{c}{WAN (ms)}\\
& (MB) & Time & \textbf{Speedup} & Time & \textbf{Speedup} \\
\hline
GAZELLE & \color{black}17.45  & 11,019.2 & \textbf{ 2.5$\times$}& \color{black}13,669.6 & \color{black}\textbf{ 1.9$\times$}\\
\color{black}DELPHI& \color{black}617 & \color{black}90,090.1&\color{black}\textbf{ 2.9$\times$} & \color{black}114,955 & \color{black}\textbf{ 2$\times$}\\
\color{black}CrypTFlow2 & \color{black}116.6  & \color{black}69,133.6 &\color{black}\textbf{ 6.5$\times$} &  \color{black}73,876.8 & \color{black}\textbf{ 4.8$\times$}\\
\cline{1-1}
\color{black}OT-based & \multirow{2}{*}{\color{black}2,108} & \multirow{2}{*}{\color{black}226,431.7} & \multirow{2}{*}{\color{black}\textbf{21$\times$}} & \multirow{2}{*}{\color{black}310,985.6} & \multirow{2}{*}{\color{black}\textbf{20$\times$}}\\
\color{black}CrypTFlow2 & & & & & \\
\hline
\hline
\multicolumn{6}{c}{Network Model: VGG}\\
\hline
GAZELLE &  \color{black}22.8 & 18,067.4 & \textbf{ 2.7$\times$}& \color{black}21,566.2 & \color{black}\textbf{ 2$\times$}\\
\color{black}DELPHI & \color{black}718.5 & \color{black}123,198.4 &\color{black}\textbf{ 2.9$\times$} & \color{black}152,176.4 & \color{black}\textbf{ 1.5$\times$}\\
\color{black}CrypTFlow2 & \color{black}150 &\color{black}97,038.9 &\color{black}\textbf{ 6$\times$} & \color{black}103,169.1 & \color{black}\textbf{ 4.6$\times$}\\
\cline{1-1}
\color{black}OT-based & \multirow{2}{*}{\color{black}5,063.7 } & \multirow{2}{*}{\color{black}340,342.9 } & \multirow{2}{*}{\color{black}\textbf{ 21$\times$}} & \multirow{2}{*}{\color{black}543,242 } & \multirow{2}{*}{\color{black}\textbf{ 24$\times$}}\\
\color{black}CrypTFlow2 & & & & & \\
\hline
\hline
\multicolumn{6}{c}{Network Model: ResNet-18}\\
\hline
GAZELLE & \color{black}54 & 42,748.3 & \textbf{ 3.2$\times$}& \color{black}51,032.7 & \color{black}\textbf{ 2.3$\times$}\\
\color{black}DELPHI & \color{black}2,033.9 & \color{black}250,618.4 &\color{black}\textbf{ 2.6$\times$} & \color{black}332,524.2 & \color{black}\textbf{ 1.9$\times$}\\
\color{black}CrypTFlow2 & \color{black}354  & \color{black}190,684.7&\color{black}\textbf{ 5.7$\times$} &  \color{black}205,146.8 & \color{black}\textbf{ 4.3$\times$}\\
\cline{1-1}
\color{black}OT-based & \multirow{2}{*}{\color{black}6,292.1 } & \multirow{2}{*}{\color{black}650,989.7 } & \multirow{2}{*}{\color{black}\textbf{ 19.5$\times$}} & \multirow{2}{*}{\color{black}903,492.6 } & \multirow{2}{*}{\color{black}\textbf{ 19$\times$}}\\
\color{black}CrypTFlow2 & & & & & \\
\hline
\hline
\multicolumn{6}{c}{Network Model: ResNet-50}\\
\hline
GAZELLE & \color{black}297.1 & 276,886.8 & \textbf{ 8.3$\times$}& \color{black}321,600.2 & \color{black}\textbf{ 4$\times$}\\
\color{black}DELPHI & \color{black}10,489 & \color{black}746,568.8 &\color{black}\textbf{ 1.7$\times$} & \color{black}1167,566.8 & \color{black}\textbf{ 1.4$\times$}\\
\color{black}CrypTFlow2 & \color{black}1,831  & \color{black}425,454.4 &\color{black}\textbf{ 4.5$\times$} & \color{black}499,429.6  & \color{black}\textbf{ 2.9$\times$}\\
\cline{1-1}
\color{black}OT-based & \multirow{2}{*}{\color{black}13,104 } & \multirow{2}{*}{\color{black}1364,463.2 } & \multirow{2}{*}{\color{black}\textbf{ 14.4$\times$}} & \multirow{2}{*}{\color{black}3307,902.6 } & \multirow{2}{*}{\color{black}\textbf{ 19$\times$}}\\
\color{black}CrypTFlow2 & & & & & \\
\hline
\hline
\multicolumn{6}{c}{Network Model: ResNet-101}\\
\hline
GAZELLE & \color{black}603.1 &486,745.2  & \textbf{ 7.7$\times$}& \color{black}577,454.9 & \color{black}\textbf{ 3.7$\times$}\\
\color{black}DELPHI & \color{black}22,199.4 & \color{black}1411,383.8 &\color{black}\textbf{ 1.7$\times$} & \color{black}2302,091.8 & \color{black}\textbf{ 1.3$\times$}\\
\color{black}CrypTFlow2 &  \color{black}3,582.8 & \color{black}777,057.4 &\color{black}\textbf{ 4.2$\times$} & \color{black}921,735.6  & \color{black}\textbf{ 2.8$\times$}\\
\cline{1-1}
\color{black}OT-based & \multirow{2}{*}{\color{black}23,857 } & \multirow{2}{*}{\color{black}2467,606.1 } & \multirow{2}{*}{\color{black}\textbf{ 13.3$\times$}} & \multirow{2}{*}{\color{black}6006,071.4 } & \multirow{2}{*}{\color{black}\textbf{ 18.2$\times$}}\\
\color{black}CrypTFlow2 & & & & & \\
\hline
\hline
\multicolumn{6}{c}{Network Model: ResNet-152}\\
\hline
GAZELLE & \color{black}873.1 & 659,833.7 & \textbf{ 7.5$\times$}& \color{black}786,587 & \color{black}\textbf{ 3.6$\times$}\\
\color{black}DELPHI & \color{black}29,433 & \color{black}1975,798.9 &\color{black}\textbf{ 1.6$\times$} & \color{black}3157,176.8 & \color{black}\textbf{ 1.3$\times$}\\
\color{black}CrypTFlow2 & \color{black}5,141  & \color{black}1065,103.4 &\color{black}\textbf{ 4.1$\times$} &  \color{black}1272,772.6& \color{black}\textbf{ 2.7$\times$}\\
\cline{1-1}
\color{black}OT-based & \multirow{2}{*}{\color{black}32,804 } & \multirow{2}{*}{\color{black}3379,188.7 } & \multirow{2}{*}{\color{black}\textbf{ 13$\times$}} & \multirow{2}{*}{\color{black}8245,124.5 } & \multirow{2}{*}{\color{black}\textbf{ 17.5$\times$}}\\
\color{black}CrypTFlow2 & & & & & \\
\hline
\hline
\end{tabular}\label{all_performance_time}
}
\end{table}

{\color{black}
In this section, we benchmark the GALA performance on a 4-layer Multi Layer Perceptron (MLP)\footnote{\color{black}The network structure is 784-128-128-10.} which is also adopted in other privacy preserving frameworks including GAZELLE, SecureML~\cite{mohassel2017secureml} and MiniONN~\cite{liu2017oblivious} as a baseline network}, as well as state-of-the-art neural network models including AlexNet~\cite{krizhevsky2012imagenet}, VGG~\cite{simonyan2014very}, ResNet-18~\cite{he2016deep}, ResNet-50~\cite{he2016deep}, ResNet-101~\cite{he2016deep}, and  ResNet-152~\cite{he2016deep}. We use MNIST dataset~\cite{mnist2020} for the MLP and CIFAR-10 dataset~\cite{CIFAR2020} for state-of-the-art networks.

Table~\ref{overall_nets_num} shows computation complexity of the proposed GALA compared with GAZELLE. We can see that GALA reduces GAZELLE's Perm by 34$\times$, 31$\times$, 30$\times$, 47$\times$, 39$\times$, and 36$\times$ for AlexNet, VGG, ResNet-18, ResNet-50, ResNet-101, and ResNet-152, respectively. The fundamental base for this speedup lies in GALA's deep optimization for HE-based linear computation. {\color{black} We also notice that GALA achieves limited reduction of Perm in MLP (from 70 to 55). This is due to the small ratio between the number of slots in the ciphertext and output dimension in each layer, i.e, $\frac{n}{n_o}$, which limits the performance gain.
{\color{black} The limited gain is also observed in Table~\ref{mlp_performance_time} which shows the system speedup of GALA over GAZELLE, CrypTFlow2, DELPHI, SecureML and MiniONN. Specifically, GALA boosts CrypTFlow2 by 2.3$\times$ in the LAN setting. SecureML also gains 2.6$\times$ in the LAN setting. Meanwhile, GALA's performance is similar to GAZELLE and MiniONN. The is due to the relatively small network size and noticeable communication overhead (i.e., the large round time in total compared with computation cost). Nevertheless, none of the competing schemes achieves a better performance than GALA.}

It is worth pointing out that the MLP network is not widely adopted in practical scenarios. On the other hand, as the state-of-the-art deep neural networks utilize large channels and small-size kernels to capture data features while the size of feature maps is large, GALA is especially effective for accelerating such large state-of-the-art network models.

Table~\ref{all_performance_time} shows the runtime of GAZELLE, DELPHI and CrypTFlow2, and the speedup of GALA on top of each.
By reducing  HE operations, especially Perm operations, GALA achieves noticeable  boost over the GAZELLE, DELPHI and CrypTFlow2 frameworks.} Specifically, the results show that GALA boosts GAZELLE by 2.5$\times$ (from 11s to 4.3s), 2.7$\times$ (from 18s to 6.5s), 3.2$\times$ (from 43s to 13s), 8.3$\times$ (from 276s to 33s), 7.7$\times$ (from 486s to 62s), and 7.5$\times$ (from 659s to 87s) in LAN setting, on AlexNet, VGG, ResNet-18, ResNet-50, ResNet-101, and ResNet-152, respectively.

CrypTFlow2 (CCS'20) is the latest framework for privacy preserved neural networks. It optimizes the nonlinear operations of DELPHI, and adopts a similar HE scheme of DELPHI for linear operations. GALA is an efficient plug-and-play module to optimize the linear operations of CrypTFlow2. As shown in the Tables~\ref{mv_performance_time} and~\ref{conv_performance_time}, GALA's optimization of linear operations can further boost CrypTFlow2 by 700$\times$ and 7.4$\times$ for matrix-vector multiplication and convolution in the LAN setting, respectively. This speedup stems from GALA's streamlined HE calculation compared with the one of CrypTFlow2. Slow-down is observed in the WAN setting, but CrypTFlow2 can still gain up to 6.5$\times$ speedup for convolution due to the computation-intensive nature for large input channels with small kernel dimensions featured in state-of-the-art network models. As for the overall system speedup, GALA can boost CrypTFlow2 by 6.5$\times$, 6$\times$, 5.7$\times$, 4.5$\times$, 4.2$\times$, and 4.1$\times$ in LAN, and by 4.8$\times$, 4.6$\times$, 4.3$\times$, 2.9$\times$, 2.8$\times$, and 2.7$\times$ in WAN, based on the aforementioned network architectures.

It might appear anti-intuitive that while CrypTFlow2 is a more recent system than DELPHI, the speedup of GALA over DELPHI is smaller than its speedup over CrypTFlow2. This is because CrypTFlow2 has optimized the nonlinear part of DELPHI, significantly reducing its runtime. As a result, the runtime of linear operations in CrypTFlow2 accounts for a very high percentage as illustrated in Table~\ref{percent_nets}. Hence CrypTFlow2 can benefit more from GALA's optimization of linear computation, resulting in a higher speedup in terms of the overall runtime. It is worth pointing out that the ability to accelerate CrypTFlow2 is highly desirable since it is the latest privacy-preserving framework.
{\color{black}Meanwhile, we also show GALA's speedup on top of the OT-based CrypTFlow2 which relies on OT to complete the linear computation. As significant communication cost, including round cost, is involved in OT, the overhead of linear computation, especially in the WAN setting, increases compared with HE-based CrypTFlow2, which results in greater speedup achieved by GALA.
}

\begin{table}[!tbp]
\footnotesize
\centering
\caption{\color{black}Percentages of linear computation in state-of-the-art neural network models.}
\begin{tabular}{c|c|c|c|c}
\hline
\hline \color{black}Networks & \color{black}GAZELLE & \color{black}DELPHI&\color{black}CrypTFlow2 & \color{black}Plaintext\\
\color{black}AlexNet & \color{black}97.7 & \color{black}76.9&\color{black}98.7 & \color{black}98.5\\
\color{black}VGG &\color{black}98.2 &\color{black}77.9 & \color{black}98.8& \color{black}98.1\\
\color{black}ResNet-18 & \color{black}98.3& \color{black}75.1& \color{black}98.6& \color{black}98.9\\
\color{black}ResNet-50 &\color{black}98.5 & \color{black}55.2& \color{black}96.8 & \color{black}97.9\\
\color{black}ResNet-101 &\color{black}98.4 & \color{black}53.2& \color{black}96.5& \color{black}98.3\\
\color{black}ResNet-152 &\color{black}98 & \color{black}52& \color{black}96.4& \color{black}98.4\\
\hline
\hline
\end{tabular}\label{percent_nets}
\end{table}

\begin{figure*}[!tbp]
\centering
\includegraphics[trim= {6.6cm 0.1cm 16cm 0cm}, clip, scale=0.40]{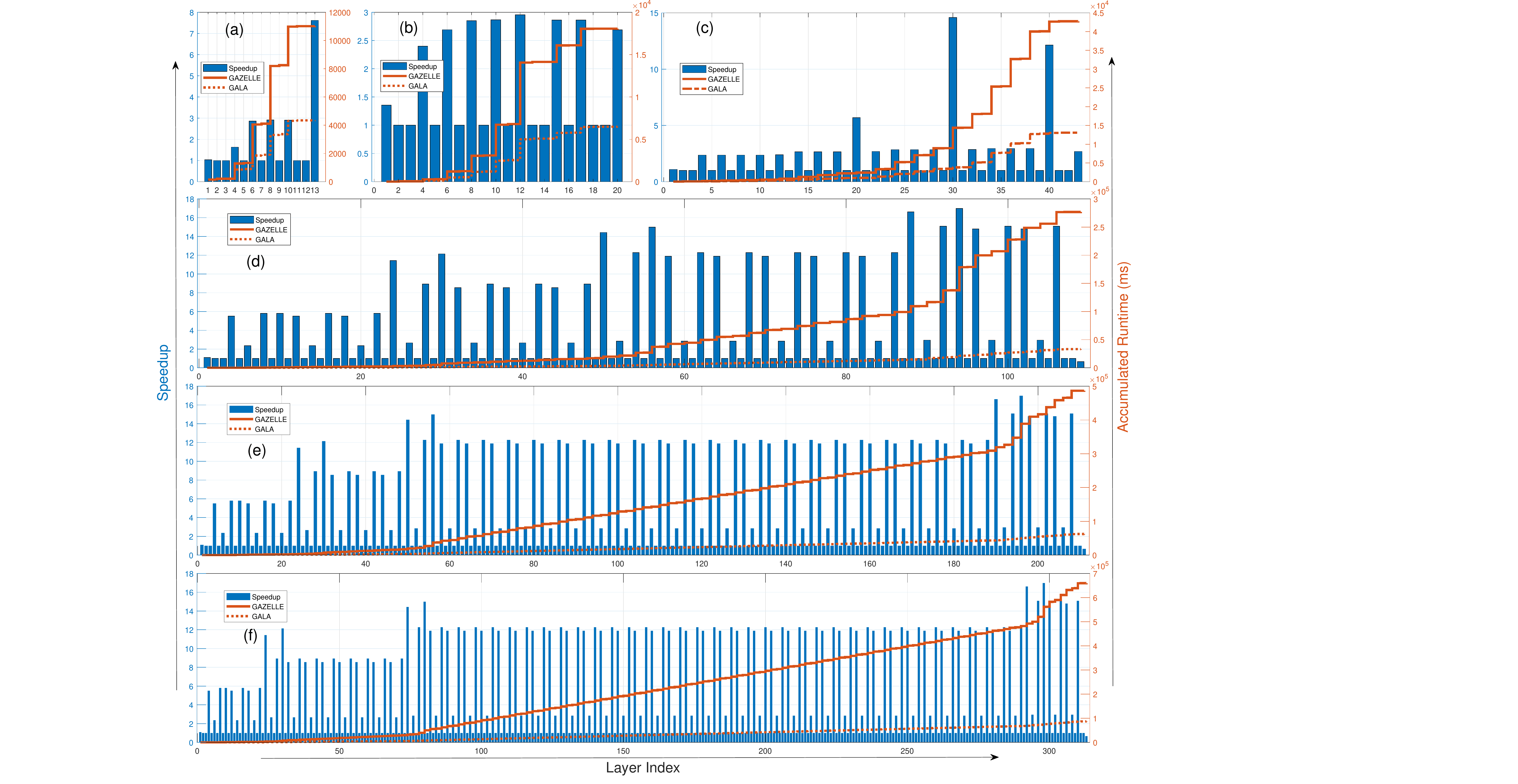}
\caption{Layer-wise accumulated runtime and GALA speedup over GAZELLA on different networks: (a) AlexNet; (b) VGG; (c) ResNet-18; (d) ResNet-50; (e) ResNet-101; (f) ResNet-152. The bar with values on the left y-axis indicates speedup, and the curve with values on the right y-axis indicates the accumulated runtime. The layers with speedup of 1 are nonlinear layers. }
\label{overall_nets}
\end{figure*}

Next we examine the runtime breakdown of different layers for those six state-of-the-art networks as shown in~Fig. \ref{overall_nets}, which allows detailed observation. Note that the layer indexing here is slightly different from the original plaintext model  for the sake of HE operations, e.g., the nonlinear activation or pooling following a convolution operation is counted as a separate layer.  The $x$-axis of each subfigure in~Fig. \ref{overall_nets} shows the layer index of a sequence of linear (convolution or matrix-vector multiplication) and nonlinear (activation or pooling) layers that constitute each network model. The $y$-axis plots the accumulated running time (milliseconds) up to a layer, and the speedup of GALA over GAZELLE in each layer.

For example, Fig. \ref{overall_nets} (a) illustrates the result for AlexNet.  The most time-consuming computation in GAZELLE is in layer ``6'', ``8'' and ``10'', which are all convolution computations. This is evidenced by the large jump  of runtime from these layers to the next layer. GALA decreases the time for these linear computations by nearly 3$\times$.
Meanwhile, the nonlinear  layers (activation/pooling) have a speedup of 1, as GALA has the same computation cost as GAZELLE in those layers. Since the nonlinear computation contributes to only a small portion of the total cost, it does not significantly affect the overall performance gain of GALA that focuses on accelerating the linear computation. Note that GALA does not benefit much in the first layer of AlexNet, i.e., the first convolution,  as the input has only three channels. However, the speedup for the following more costly convolutions allows GALA to effectively reduce the overall cost.
A similar observation can be seen from the result on VGG. As for the four ResNets frameworks, the most significant performance gain stems from the convolution with 1$\times$1 kernels. As ResNets repeat the blocks with multiple 1$\times$1 convolution kernels, GALA effectively accelerates this type of convolution due to its deeply optimized linear computation mechanism (see details in Sec.~\ref{system:conv}), thus reducing the overall runtime. {\color{black}Similar trend is observed for DELPHI and CrypTFlow2}.

{\color{black}
It is also worth mentioning that GALA focuses on optimizing the HE-based linear operations only and can be integrated into a baseline model (such as GAZELLE, CryptFlow2, or DELPHI). The proposed approach does not introduce approximation. Hence it does not result in any accuracy loss compared to the baseline privacy preserved model. Furthermore, compared with the original plaintext model, the only possible accuracy loss in GALA comes from the quantification of floating point numbers to fixed point numbers in the HE operations. Such quantification is indispensable in all HE-based frameworks including CryptFlow2. From our experiments, the model accuracy loss due to quantification is negligible, as shown in Table~\ref{nets_accuracy}.
}

\begin{table}[!tbp]
\small
\centering
\caption{\color{black}Accuracy with floating and fixed point in state-of-the-art neural network models. Top-1 accuracy: only the prediction with the highest probability is a true label; Top-5 accuracy: any one of the five model predictions with higher probability is a true label.}
\begin{tabular}{c|c|c|c|c}
\hline
\hline
\multirow{2}{*}{\color{black}Network Models} & \multicolumn{2}{c|}{\color{black}Floating-point } & \multicolumn{2}{c}{\color{black}Fix-point }\\
\cline{2-5}
 & \color{black}Top1 & \color{black}Top5 & \color{black}Top1 & \color{black}Top5\\
 \hline
 \color{black}AlexNet &\color{black}78.89\% &\color{black}97.32\% &\color{black}78.43\% &\color{black} 97.26\% \\
 \color{black}VGG &\color{black}92.09\% &\color{black}99.72\% &\color{black}92.05\% &\color{black}99.68\% \\
 \color{black}ResNet-18 &\color{black}93.33\% &\color{black}99.82\% &\color{black}93.21\% &\color{black} 99.81\%\\
 \color{black}ResNet-50 &\color{black}93.86\% &\color{black}99.85\% &\color{black} 93.86\% &\color{black}99.84\% \\
 \color{black}ResNet-101 &\color{black}94.16\% &\color{black}99.79\% &\color{black}94.12\% &\color{black}99.79\% \\
 \color{black}ResNet-152 &\color{black}94.23\% &\color{black}99.81\% &\color{black}94.15\% &\color{black}99.79\% \\
\hline
\hline
\end{tabular}\label{nets_accuracy}
\end{table}

%% file: conclusion.tex
\section{Conclusion and Further Discussions}\label{conclusion}
This paper has focused on a deep optimization on the HE-based linear computation in privacy-preserved neural networks. It aims to minimize the Perm operations, thus substantially reducing the overall computation time. To this end, we have
proposed {\em GALA: \underline{G}reedy comput\underline{A}tion for \underline{L}inear \underline{A}lgebra}, which views the HE-based linear computation as a series of Homomorphic Add, Mult and Perm operations and chooses the least expensive operation in each linear computation step to reduce the overall cost. GALA has made the following contributions: (1) It has introduced a row-wise weight matrix encoding with combined share generation (i.e., row-encoding-share-RaS (Rotated and Sum)) to reduce the Perm operations for dot product. (2) It has designed a first-Add-second-Perm approach (named {\em kernel grouping}) to reduce the Perm operations for convolution.
{\color{black}As such, GALA efficiently reduces the cost for the HE-based linear computation, which is a critical building block in almost all of the recent frameworks for privacy-preserved neural networks, including GAZELLE, DELPHI, and CrypTFlow2. With its deep optimization of the HE-based linear computation, GALA can be a plug-and-play module integrated into these systems to further boost their efficiency. Our experiments show that GALA achieves a significant speedup up to 700$\times$ for the dot product and 14$\times$ for the convolution computation under different data dimensions.}
Meanwhile, GALA demonstrates an encouraging runtime boost by 2.5$\times$, 2.7$\times$, 3.2$\times$, 8.3$\times$, 7.7$\times$, and 7.5$\times$ over GAZELLE
and 6.5$\times$, 6$\times$, 5.7$\times$, 4.5$\times$, 4.2$\times$, and 4.1$\times$ over CrypTFlow2, on AlexNet, VGG, ResNet-18, ResNet-50, ResNet-101, and ResNet-152, respectively.

It is worth pointing out that even with the significant progress toward privacy preserved machine learning in recent years (including this work), there still exists a large performance gap between the plaintext system (generally below a second) and the privacy preserved system (ranging from seconds to hundreds of second). Nevertheless, it is still promising to attain the long-term goal for practical implementation of privacy preserved machine learning. First, the privacy preserved machine learning system is to be deployed on clouds with abundant computation power.
Hence, even though it takes significantly more time than the plaintext system on the same local hardware, running it on clouds with parallel computing infrastructure can significantly reduce the gap. Second, the research efforts on the in-depth optimization of the privacy-preserved computation further help to close the runtime gap. Altogether, the combination of advanced algorithms and cloud computation resources may enable the privacy-preserved system to achieve a response time well suited for some practical applications in the near future.